\documentclass[twocolumn,usenatbib,useAMS]{mnras}  

\usepackage{dsfont}
\usepackage{amsmath}
\usepackage{amssymb}
\usepackage{graphicx}
\usepackage{verbatim}
\usepackage{natbib}
\usepackage{eucal}
\usepackage{calligra}
\usepackage{soul}
\usepackage{amsfonts}
\usepackage{color}
\usepackage[normalem]{ulem}
\usepackage[T1]{fontenc}

\usepackage{times}
\usepackage{epsfig}
\usepackage[usenames,dvipsnames]{xcolor}
\usepackage{url}

\DeclareGraphicsExtensions{.png,.pdf}
\DeclareMathAlphabet{\mathscr}{OT1}{pzc}{m}{it}

\newcommand{\be}{\begin{equation}}
\newcommand{\ee}{\end{equation}}
\newcommand{\bes}{\begin{equation*}}
\newcommand{\ees}{\end{equation*}}
\newcommand{\bea}{\begin{eqnarray}}
\newcommand{\eea}{\end{eqnarray}}
\newcommand{\beas}{\begin{eqnarray*}}
\newcommand{\eeas}{\end{eqnarray*}}

\def\[{\begin{equation}}
\def\]{\end{equation}}

\def\mpcoh{{\,h^{-1}\,\rm Mpc}}
\def\v#1{{\bf #1}}
\renewcommand{\vec}[1]{{\boldsymbol{#1}}}

\def\citejap#1{\citeauthor{#1}\ \citeyear{#1}}

\def\m@th{\mathsurround=0pt }
\def\eqalign#1{\null\,\vcenter{\openup1\jot \m@th
 \ialign{\strut\hfil$\displaystyle{##}$&$\displaystyle{{}##}$\hfil
 \crcr#1\crcr}}\,}
\def\beq{\begin{equation}}
\def\eeq{\end{equation}}
\def\gs{\mathrel{\lower0.6ex\hbox{$\buildrel {\textstyle >}
 \over {\scriptstyle \sim}$}}}
\def\ls{\mathrel{\lower0.6ex\hbox{$\buildrel {\textstyle <}
 \over {\scriptstyle \sim}$}}}
\def\japsc{\scriptscriptstyle\rm}
\def\omit#1{}
\def\fg{f}

\include{aas_journals}

\usepackage{soul,xcolor}

\begin{document}
\title[Detection of cosmological dipoles aligned with transverse peculiar velocities]
  {Detection of cosmological dipoles aligned with transverse peculiar velocities}

\author[Y.-C. Cai et al.]{
Yan-Chuan Cai$^{1}$\thanks{E-mail: cai@roe.ac.uk},
John A. Peacock$^{1}$,
Anna de Graaff$^{2}$,
Shadab Alam$^{3}$
\\
$^{1}$Institute for Astronomy, University of Edinburgh, Royal Observatory Edinburgh, Blackford Hill, Edinburgh EH9 3HJ, UK\\
$^2$Max-Planck-Institut f\"ur Astronomie, K\"onigstuhl 17, D-69117 Heidelberg, Germany \\
$^3$Department of Theoretical Physics, Tata Institute of Fundamental Research, Homi Bhabha Road, Mumbai 400005, India
}

\setstcolor{red}

\maketitle
\begin{abstract}
Peculiar velocities encode rich cosmological information, but their transverse components are hard to measure. Here, we present the first observations of a novel effect of transverse velocities: the dipole signatures that they imprint on the Cosmic Microwave Background. 
The peculiar velocity field points towards gravitational wells and away from potential hills, reflecting a large-scale dipole in the gravitational potential, coherent over hundreds of Mpc. Analogous dipoles will also exist in all other fields that correlate with the potential. These dipoles are readily observed in projection on the CMB sky via gravitational lensing and the integrated Sachs-Wolfe (ISW) effect -- both of which correlate with transverse peculiar velocities. The large-scale ISW dipole is distinct from the small-scale moving lens effect, which has a dipole of the opposite sign. We provide a unified framework for analysing these velocity-related dipoles and demonstrate how stacking can extract the signal from sky maps of galaxy properties, CMB temperature, and lensing. We show that the CMB dipole signal is independent of galaxy bias, and orthogonal to the usual direction-averaged correlation function, so this new observable provides additional cosmological information. We present the first detections of the dipole signal in (i) galaxy density; (ii) CMB lensing convergence; and (iii) CMB temperature -- interpreted as the ISW effect -- using galaxies from the SDSS-III BOSS survey and CMB maps from {\it Planck}. We show that the observed signals are consistent with $\Lambda$CDM predictions, and use the combined lensing and ISW results to set limits on linearised models of modified gravity.  
\end{abstract}


\section{Introduction}

Just prior to the epoch of recombination, at a redshift of approximately $z=3500$, our universe entered the era of domination by nonrelativistic matter. After the final scattering of the Cosmic Microwave Background (CMB), at $z\simeq 1100$, the 6D phase-space distribution of matter, [$\rho(\vec{r})$, $\vec{v(r)}$], provides the main source of information regarding physical cosmology. The 3D density field can be inferred primarily by using galaxies as proxies of the matter fluctuations, although the density field can also be observed in 2D projection through gravitational lensing. But it is more challenging to measure peculiar velocities, the deviations from uniform expansion that are inevitably associated with the growth of structure.
The radial components of these velocities can be detected because they add a Doppler modification to the ideal cosmological redshift:
\beq
(1+z) \to (1+z) (1+v_\parallel/c)
\eeq
(assuming $v_\parallel$ to be nonrelativistic). If we have a means of estimating comoving distances directly through some standard-candle relation, then $v_\parallel$ can be estimated via
\beq
v_\parallel = {H(z)\over 1+z}\,
\left(D_z - D_{\rm direct}
\right),
\eeq
where $D_z$ is the standard homogeneous comoving distance-redshift relation based on the observed redshift, and $H(z)$ is the Hubble parameter at redshift $z$ \citep[e.g.][]{2016AJ....152...50T, 2022MNRAS.515..953H}. This requires us to assume a cosmological model in order to calculate $H(z)$ and $D_z$; but in practice this introduces little uncertainty because the fractional errors in $D_{\rm direct}$ are generally around 20\%, so useful direct estimates of $v_\parallel$ can only be obtained for galaxies at $z\ls0.05$.

There are two major ways of inferring the $v_{\parallel}$ field indirectly:
One is through the impact of $v_{\parallel}$ on the apparent density field traced out by galaxies via the effect of redshift-space distortions \citep{Kaiser1987}. Peculiar velocities of galaxies perturb their distribution along the line of sight (LOS) while leaving the transverse distribution intact. This causes a statistical anisotropy in the distribution of galaxies, and analysing this signature allows us to extract information about the LOS peculiar velocities (e.g. \citejap{2001Natur.410..169P}, \citejap{2009MNRAS.393..297P}, \citejap{2017MNRAS.470.2617A}).
A second probe is given by  the impact on the CMB of $v_{\parallel}$ of free electrons associated with galaxies. Scattering by these electrons increases/decreases the line-of-sight CMB temperature if the electrons are moving towards/away from the observer; this is known as the kinetic Sunyaev--Zeldovich (kSZ) effect \citep{Sunyaev_Zeldovich1972,Sunyaev1980}. In addition, line-of-sight peculiar velocities of galaxies may also be detected via angular redshift fluctuations, as discussed in \citet{Hernadez2021}.

Measurements of the transverse velocity, $v_{\perp}$,  are more difficult. For very nearby objects we can use high precision astrometry to monitor proper motions on the sky and thus infer $v_{\perp}$ directly, but this is barely feasible even for the closest galaxies (\citejap{M31_vperp}; see also \citejap{Hall2019}). Micro-lensing of quasars by galaxies has also been explored as a probe of transverse velocity \citep{Mediavilla2016}, but somewhat strong modelling assumptions are needed. The transverse velocities can be estimated indirectly over cosmological volumes only by making additional assumptions. A common approach is to assume that the large-scale component of the velocity field is associated with fluctuations in the growing mode of gravitational instability. In that case, the velocity field should be irrotational and generated as the gradient of a velocity potential. This potential can be calculated by radial integration of $v_\parallel$, and then $v_{\perp}$ can be reconstructed. This programme was influential in the 1990s (e.g. \citejap{POTENT}), but was limited by the depth of peculiar-velocity surveys. For a still less direct estimate, one can appeal to the continuity equation applied to galaxies. If biased galaxy density fluctuations are in the linear regime, then
\beq
{\bf \nabla\cdot u}=-H(z) \fg(z) \delta_g/b,
\eeq
where $\bf u$ is the comoving peculiar velocity field, ${\bf u =v}/a(t)$ where $a(t)$ is the cosmic scale factor, $\delta_g$ is the galaxy fractional density fluctuation, $b$ is the bias parameter, and $\fg =d\ln\delta/d\ln a$
is the fluctuation growth rate. In the growing mode, the velocity associated with a given Fourier mode is parallel to its wavevector, so that this equation can be inverted in Fourier space. Thus the spatial distribution of galaxies can be used to predict the peculiar velocity field, within an overall scaling factor. This method is now very commonly applied, since it is of interest to have the velocity field and associated displacement field in order to apply `reconstruction' techniques to remove the effects of nonlinear evolution on the signature of Baryon Acoustic Oscillations (BAO); see e.g. \cite{2012MNRAS.427.2132P}.
 
Alternatively, we may look for indirect observational signatures of transverse motions. There are at least three physical effects associated with transverse motions of matter on cosmological scales: the relativistic transverse Doppler effect, which is quadratic in the velocity \citep{Zhao2013,Kaiser2013}; the polarised kSZ signal, which is also proportional to $v^2_{\perp}$ \citep{Sunyaev1980}; and the moving gravitational-lens effect \citep{Birkinshaw1983}, which is linear in the velocity. So far, none of these effects have been detected. Our focus here will be relevant to the third effect, which operates on larger scales. 
One way of understanding the moving lens effect is as a Doppler shift, where the lens deflection induces an effective LOS velocity of $v_\perp\alpha\,\theta$, where $\alpha$ is the lensing deflection angle, and $\theta$ is the angular offset on the sky in the direction of the transverse velocity vector.
The moving lens thus induces a dipole of CMB temperature decrement and increment (cold and hot spot) aligned with the transverse velocity vector -- as explained in more detail in Section \ref{sec:moving_lens}. 
One of the initial motivations for the present work was to seek a detection of this dipole. 

A more general way of thinking about the moving gravitational lens is as a specific contribution to the Integrated Sachs-Wolfe (ISW) effect \citep{ISW1967, 1990ApJ...355L...5M}. 
Here we consider the gravitational effect on (CMB) photons traversing the time-varying large-scale structure. If the gravitational potential $\Phi(\vec r)$ varies with time, the energy of the photon will be altered by an amount proportional to the change of the potential along its trajectory. The temperature perturbation is
\begin{equation}
 \frac{\Delta T}{T_{\japsc CMB}} = \frac{2}{c^2}\int \dot\Phi\, dt, 
 \end{equation}
where $dt$ is an element of cosmic time along the photon trajectory, 
viewed as a positive quantity. The opposite sign is given by \citet{Rubino-Martin2004}, but this traces to \citet{Cooray2002}, whose expression (267) defines $\dot\Phi$ as the derivative with respect to look-back time, which introduces a negative sign. 
In the case of the moving lens the local gravitational potential on the leading side of the lens is deepening due to the approach of the gravitational potential well, whereas on the trailing side the local gravitational potential is becoming shallower. The resulting temperature dipole alignment with the direction of transverse motion was discussed in \cite{Cai2010} -- see Figs~3 \& 7 of their paper -- and interpreted as a nonlinear ISW phenomenon i.e. the Rees-Sciama effect \citep{Rees1968}. 

The general ISW perturbation is rather small, $\Delta T/T_{\japsc CMB} \sim 10^{-6}$, and subdominant to the primordial CMB. This signal has been extracted by using information about galaxy density fluctuations: either cross-correlating those with the CMB \citep[e.g.][]{Crittenden1996,Fosalba2003, Giannantonio2008, Ho2008, PlanckISW2014}, or stacking CMB maps at the locations of superstructures \citep[e.g.][]{Granett2008, Ilic2013, PlanckISW2014, Cai2014, Nadathur2016, Kovacs2017, Hang2021b, Dong2021}. In either case, the significance of detection is only modest; although the lack of large ISW effects is a strong constraint on non-standard cosmological models. A more general expression for the temperature perturbation is
\begin{equation}
 \frac{\Delta T}{T_{\japsc CMB}} = \int (\dot\psi+\dot\phi)\, dt, 
 \end{equation}
where we assume the
Newtonian gauge in which the time and spatial parts of the metric are perturbed by the dimensionless Bardeen potentials
$\psi$ and $\phi$ respectively. Gravitational redshift depends purely on $\psi$, while both the ISW and gravitational lensing effects probe the sum of the potentials, $\psi + \phi$ \citep{Yoo2009,Bonvin2011,Challinor2011, Kaiser2013}. 
In the Newtonian limit, without any anisotropic stress, $\psi = \phi =\Phi/c^2$ and the ISW effect depends only on the Newtonian gravitational potential. In modified gravity models there may be a gravitational slip, i.e. $\psi\neq\phi$; the combination of the above effects provides a means of testing such non-Einstein theories \citep{Clifton2012, Simpson2013}, but in making ISW predictions we shall adopt the standard assumption of zero slip.

For the present paper, the example of the moving-lens CMB dipole suggests a new way of probing the ISW effect. This is expected to show a dipole that aligns with the transverse peculiar velocity in the linear regime, rather than just as a result of the nonlinear Rees-Sciama effect -- and this linear effect dominates at large scales. The same principle applies to related gravitational effects such as gravitational lensing.
In the linear regime, $\phi(\vec r)$ is constant while the universe is matter dominated, but these primordial fluctuations decay when the expansion becomes dominated by dark energy or by curvature (but here we assume flat models). Gradients in the potential dictate the peculiar gravitational acceleration, which is parallel to the peculiar velocity in the linear regime. The local transverse velocity therefore points in the direction of a large-scale potential well, and points away from a potential hill. It is then to be expected that the potential field will exhibit a large-scale dipole that aligns with the peculiar velocity. Furthermore, there will be aligned dipoles in all quantities that are linearly related to potential, including density, ISW and gravitational lensing signals. We will give a unified treatment of these dipoles and present attempts to detect them observationally. In so doing, we will be testing a novel signature of the standard picture of gravitational instability, and we can also hope that this approach of bringing in the external velocity data may help improve the detection of weak effects such as the ISW signal. Some elements of this idea have been investigated by others: in particular the papers by \cite{hotinli2023} and \cite{hotinli2024} discuss the relation between transverse velocities and the projected dipole in the thermal Sunyaev--Zeldovich effect. \cite{Beheshti2024} have made detailed predictions of the moving-lens effect. However, none of these papers has dealt with the large-scale linear effects that are the principal focus of the present paper.

\begin{figure*}
\begin{center}
\vspace{-0.0cm}
\scalebox{1.0}{
	\includegraphics[width=0.9\textwidth]{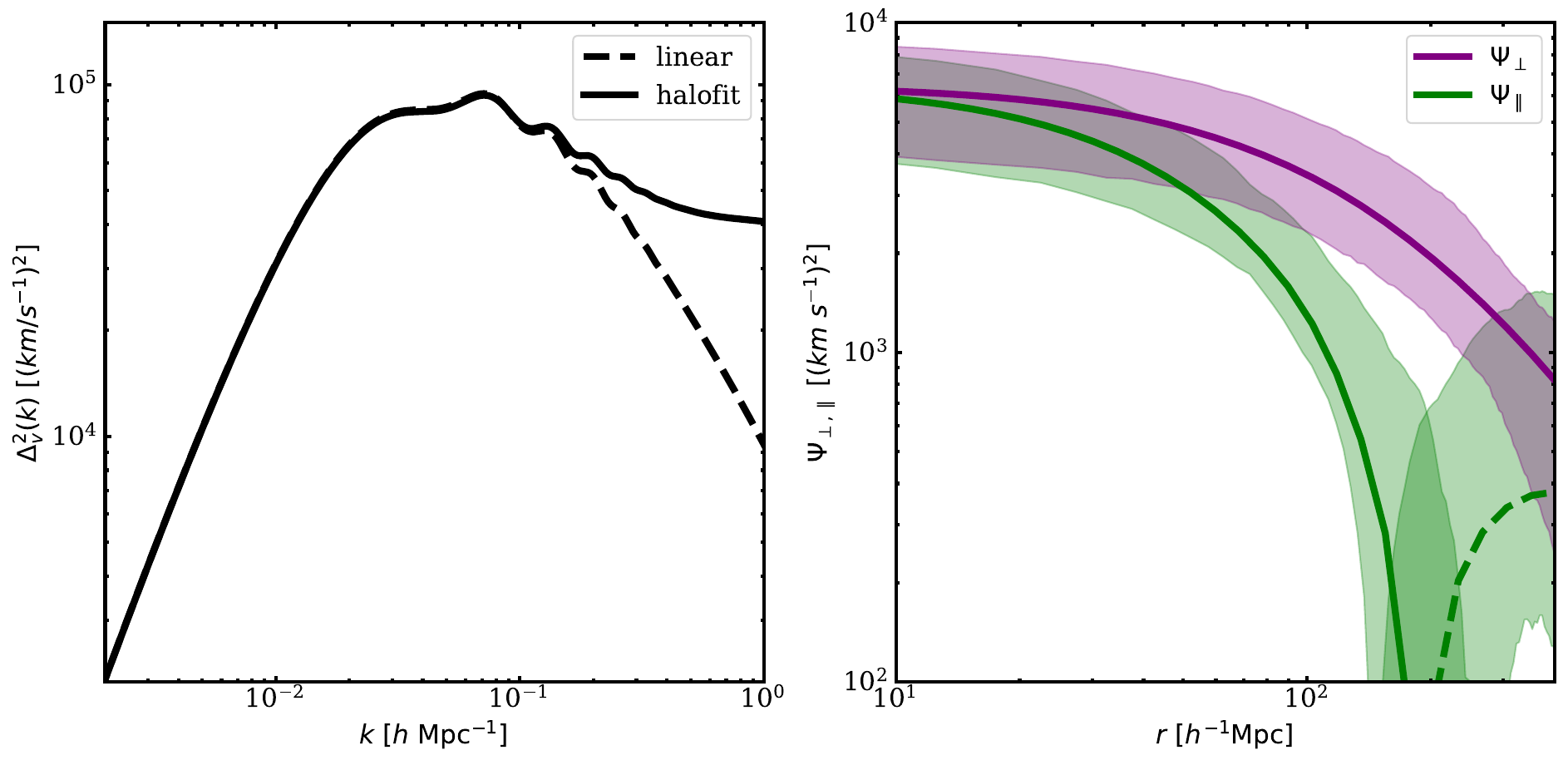}}
    \caption{Left: dimensionless velocity power spectrum for a $z=0.55$ flat $\Lambda$CDM universe with $\Omega_m=0.292$. $\Delta^2_v$ in the left panel is the 3D velocity variance per $\ln k$: $(aH\fg)^2 k P/2\pi^2$. Solid / dashed lines are results from {\sc halofit} \citep{Smith2003, halofit2} / linear theory. Right: the parallel ($\Psi_{\parallel}$) and perpendicular ($\Psi_{\perp}$) components of the velocity correlation as a function of comoving separation $r$. Coloured bands are measurements from N-body simulations, discussed in Appendix \ref{appA}. The widths of the bands represent $\pm 1$-$\sigma$ errors for a volume of (1380~$h^{-1}$Mpc)$^{3}$.
    Solid lines are predictions from linear theory (Eq.~\ref{eq:psi}); the dashed line indicates negative values. 
    }
    \label{v-correlation-1D}
\end{center}
\end{figure*}

The practical means of achieving this observational search is `stacking with rotation'. We take the direction of the transverse velocity at a given point as known and expect that there will be a dipole in some scalar quantity (e.g. CMB temperature) about that point,
aligned with the velocity. We can therefore average the maps of interest around a number of stacking centres (most simply, the galaxies in a given catalogue) by rotating each map so that ${\bf v}_\perp$ points in a common direction, and then averaging these rotated maps.
However, there are two major challenges in achieving this: 
(i) there is a substantial cosmic variance in the dipole signal, so that even in ideal data we need to average over a large number of map centres;
(ii) there may well be additional signals in the data. The amplitudes of the temperature dipoles are expected to be small ($\Delta T/T \sim 10^{-6}$), and so they risk being overwhelmed by the primordial CMB temperature fluctuations. Both these effects need to be understood statistically in order to quantify the precision of any detection.
 
We begin in Section~\ref{sec:vcorrelations} by discussing the statistical properties of the coherent peculiar velocity field. This diverges from low density regions and converges towards high density regions, giving rise to a velocity dipole resembling the field of an electric dipole on scales of several hundred Mpc. We then show in Section~\ref{sec:dipole_prediction} that the velocity dipole traces the gradients of its underlying gravitational potential, so that it is coupled to all gravitational effects, particularly ISW \& gravitational lensing. In Section~\ref{sec:ISW_Lensing_Dipoles}, we make detailed predictions for these signals. In Section~\ref{sec:observation} we discuss the peculiar velocity reconstruction to be used, which we take from SDSS-BOSS, and we verify that this correlates as expected with the projected galaxy density. In Section~\ref{sec:results} we then present the attempted detection of the gravitational dipoles in ISW and gravitational lensing, using data from {\it Planck\/}. Section~\ref{sec:interpretation} gives the interpretation of these results in terms of the growth of density fluctuations and constraints on models of gravity.
We give a summary and discussion in Section~\ref{sec:summary}.

Generally we assume a flat $\Lambda$CDM cosmology, with parameters taken from the most recent Planck PR4 release \citep{Planck_PR4}. Small deviations from this choice are sometimes needed when dealing e.g. with public simulation data, but such differences are generally insignificant for the present purpose.
 
\section{Velocity fields and correlated dipoles}

\subsection{Velocity correlations}
\label{sec:vcorrelations}

The theory of correlated velocity fields is presented in a usefully comprehensive form by \citet{Gorski1988A}. 
In what follows, note that it is normal to describe lengths using comoving units, whereas with velocities it is common to require the proper cosmological peculiar velocity (which generates Doppler shifts). We will therefore use $v$ to denote proper velocities, as distinct from $u$ which will denote a comoving velocity: $v=a(t)u$, where $a(t)$ is the cosmological scale factor.

\begin{figure*}
\begin{center}
\scalebox{1.0}{ \includegraphics[width=1.0\textwidth]{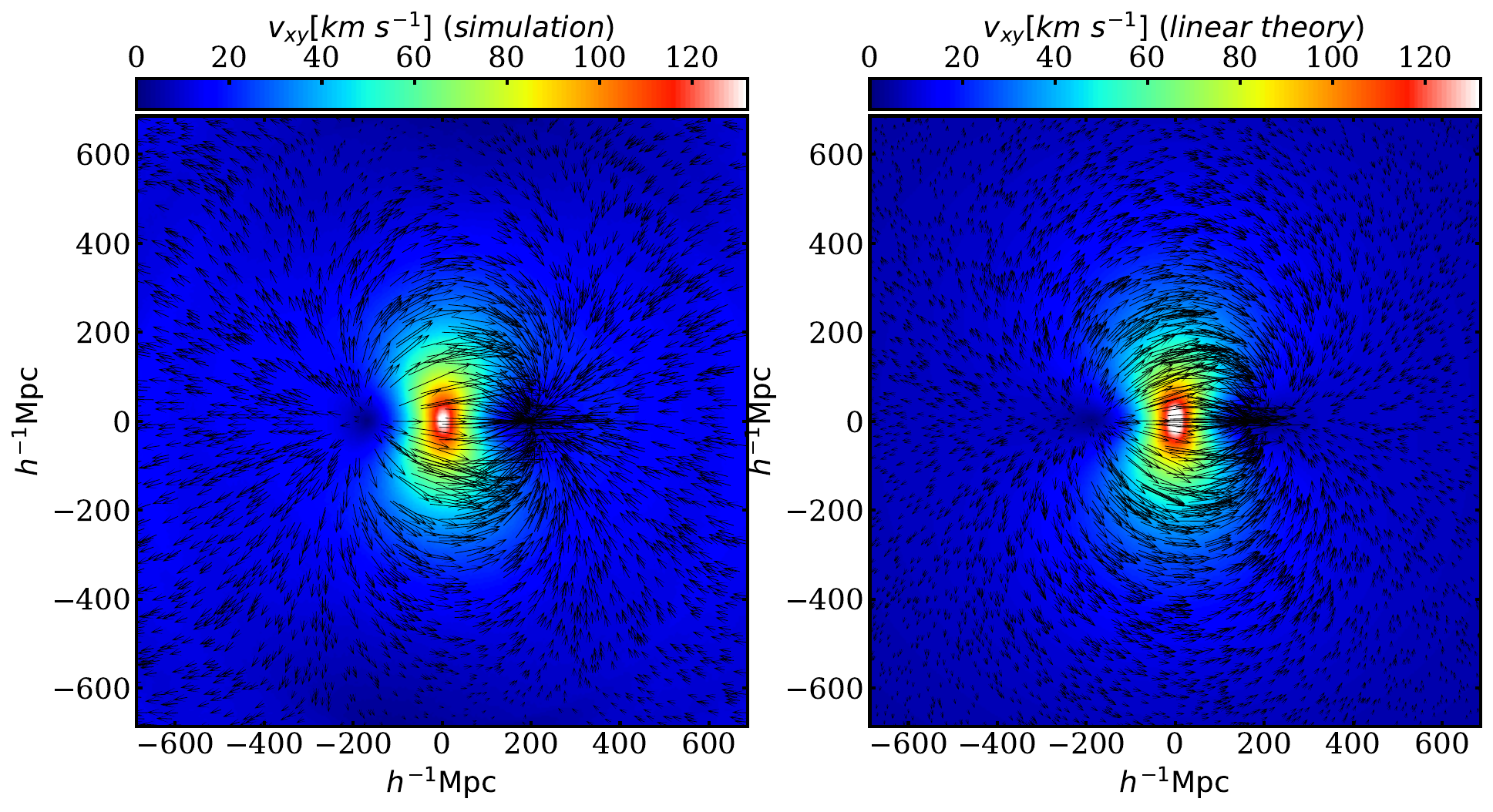}}
 \scalebox{1.0}{\includegraphics[width=1.0\textwidth]{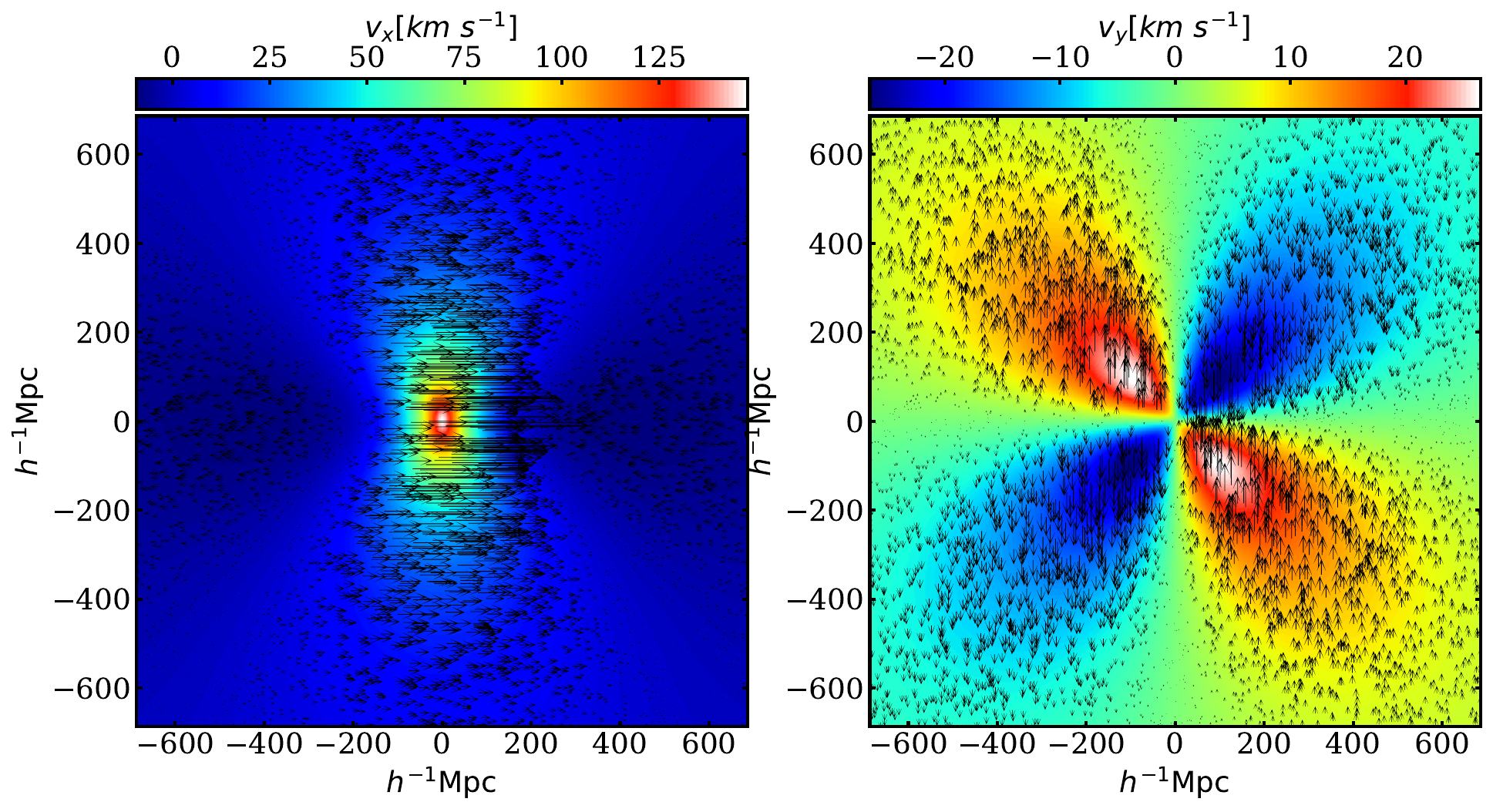}}
		    \caption{Top panels: $z=0.55$ stacked 2D velocity field predicted from linear theory as described in Section \ref{sec:vcorrelations} (right) and the measurement from simulations, as discussed in Appendix \ref{appA} (left). 
            Colours represents the amplitude of velocities, and arrows indicate velocity vectors. We can see that the velocity vectors at the centre of the plots correctly indicate that matter flows from low density to high density regions. Bottom-left and bottom-right panels show the $x$ and $y$-component of the velocity field according to linear theory, revealing a dipole and a quadrupole, respectively.}
\label{v-correlation-2D}
		\end{center}
\end{figure*}
The covariance of velocity components at comoving separation $\bf r$ can be written as
\begin{equation}
\langle v_i v’_j \rangle \equiv \Psi_{ij} = \Psi_\perp(r)\, \delta_{ij} + [\Psi_\parallel(r) -\Psi_\perp(r)]\, r_i r_j/r^2,
\end{equation}
where the two velocity correlation functions are given in linear theory by
\begin{equation}
\label{eq:psi}
\Psi(r) = a^2H(z)^2\fg^2(z) \int \Delta^2(k,z)\, k^{-2}\, d\ln k\, K(kr),
\end{equation}
with the kernels being respectively
\begin{eqnarray}
K_\perp(x) &=& [\sin(x)-x\cos(x)]/x^3; \\
K_\parallel(x) &=& \sin(x)/x - 2[\sin(x)-x\cos(x)]/x^3.
\end{eqnarray}
The subscripts $\perp$ and $\parallel$ represent components of pairwise velocities that are perpendicular and parallel to the position vector that connects the pair. The right-hand panel of Fig.~\ref{v-correlation-1D} compares predictions from linear theory with N-body simulations. We see that $\Psi_\parallel$ crosses zero at about $200\mpcoh$, whereas $\Psi_\perp$ is always positive. Both correlation functions become equal at small separations, so that $\Psi_{ij}(0)=\Psi_\perp(0)\delta_{ij}$. 

If we define the velocity correlation $\psi_{ij}(r) = \Psi_{ij}(r)/\Psi_{\perp}(0)$, then the expectation of the velocity at $\bf r$ conditional on the velocity at the origin is
\begin{equation}
\langle v_i({\bf r}) \rangle = \psi_{ij}\, v_j(0).
\end{equation}
If we apply this to a coordinate system where $\bf v(0)$ lies along the $x$ axis, then the 2D velocity field as a function of $x$ and $y$ is
\begin{equation}
\label{eq:v2D}
\langle {\bf v} \rangle = |{\bf v(0)}|\, {1\over r^2}
\begin{bmatrix}
\psi_\perp y^2 + \psi_\parallel x^2 \\
(\psi_\parallel-\psi_\perp)\, xy
\end{bmatrix}\, .
\end{equation}
Thus $v_x$ is symmetric while $v_y$ is antisymmetric. 
Note that an unobserved non-zero $z$ component of velocity does not affect this result: if the two points concerned are separated only in the plane of the sky, then the structure of the correlation tensor says that $\psi_{xz}=\psi_{yz}=0$, and the radial velocity does not enter. Thus $\bf v(0)$ can be taken as being just the 2D velocity components in the plane of the sky, i.e. the transverse velocity $\bf v_\perp(0)$. The amplitude of this velocity can be estimated using linear theory: 
\[\langle|{\bf v_\perp(0)}|\rangle=\sqrt{(\pi/6)\langle v^2\rangle_{\rm 3D}}.
\] 

\subsubsection{Velocities averaged in a slab}
For the impact of peculiar velocities on the CMB and other gravitational effects, we are interested in the statistics of the transverse components. Because of statistical isotropy, the general 3D form of the velocity correlation tensor applies to this case immediately. So, for example, $\Psi_\parallel(r)$ is the autocorrelation of $v_x$ for points separated in $x$ by a distance $r$, and $\Psi_\perp(r)$ is the autocorrelation of $v_y$ for points with the same $x$ separation. In other words, it suffices to have the velocity field in an infinitesimal slice at constant $z$ and to ignore $v_z$, as discussed earlier. But sometimes we may wish to produce a 2D velocity field by averaging $v_x$ and $v_y$ through a slab of finite thickness. Let $W(k_z)$ be the window function corresponding to this slice -- e.g. $W=\sin(k_zL/2)/(k_zL/2)$ for a uniformly weighted slice of depth $L$. The 2D velocity correlations for this case are
\begin{eqnarray}
\label{eq:psiproj}
\Psi(r) &=& {1\over 2}\,a^2H(z)^2\fg^2(z) \int_{-\infty}^\infty
\int_0^\infty \Delta^2(k,z)\, k^{-7}\, \nonumber \\
&&  \times K(k_\perp r)\, k_\perp^3 dk_\perp\; |W|^2(k_z)\, dk_z,
\end{eqnarray}
where the transverse wavenumber is $k_\perp\equiv(k^2-k_z^2)^{1/2}$,
the kernels are respectively
\begin{equation}
K_\perp(x) = J_1(x)/x; \ \ \
K_\parallel(x) = J_0(x)-J_1(x)/x,
\end{equation}
where $J_0$ and $J_1$ are Bessel functions, and $\Delta^2(k,z)\equiv k^3P(k,z)/(2\pi^2)$ is the dimensionless power spectrum. For a simulation set up in a cubic box, there are lower and upper bounds for $k$ e.g., $k_{\rm min}=2\pi/L_{\rm box}$ and $k_{\rm max}=k_{\rm Ny}$, the Nyquist frequency. 

\subsubsection{Predicted and simulated velocity fields}

We can now compare these analytical calculations with measurements from N-body simulations. Fig.~\ref{v-correlation-2D} shows the mean velocity field as computed in linear theory and as measured in N-body simulations (See Appendix \ref{appA} for details). We can see a dipole pattern in the stacked velocity field, which reveals the nature of the bulk flow: matter diverges from low-density regions (voids: positive values of gravitational potentials) and converges towards high-density regions (superclusters: negative values of gravitational potentials). This is evident when we overlay the underlying gravitational potential field, which is linearly proportional to the temperature fluctuation from the ISW effect (Fig.~\ref{Tisw-2D}), and also when we show the matter density fluctuation field (Fig.~\ref{kappa-2D}).
The dipole pattern of the velocity field closely resembles the electric field generated by an electric dipole, with the centres of convergence/divergence playing the role of positive/ negative electric charges. This resemblance is no coincidence: both the gravitational and electric potentials obey the Poisson equation, and the linear peculiar velocity is proportional to the gradient of the gravitational potential.

The coherence of the dipole extends to several hundred Mpc, but the pattern is dictated simply by the direction of the transverse velocity in the centre, averaged over a relatively small volume e.g. $\sim$\,$10\mpcoh$ or less. This reflects the fact that the velocity field in configuration space receives contributions from the very large-scale (low-$k$) perturbation modes. It is also clear that the analytical prediction from linear theory agrees very well with measurements from N-body simulations down to scales of $\sim$10 Mpc. At even smaller scales, the exact velocity field will deviate from linear theory; but we will not be concerned with such scales in this paper and will generally be content to assume a linear model for the velocities.

To appreciate the connection between the mathematical form of the dipole from Eq.~(\ref{eq:v2D}) and its physical appearance, we decompose the dipole into two orthogonal components, also shown in Fig.~\ref{v-correlation-2D}. For the $v_x$ component (left), we can see that matter flows away horizontally from the divergence point at approximately $(-200,0)\mpcoh$, and flows into the convergence centre at $(200,0)\mpcoh$. The same pattern is seen for the $v_y$ component, except that it occurs along the vertical direction. This forms a quadrupolar pattern, expected from Eq.~(\ref{eq:v2D}). These features are all in good agreement with N-body simulations, which are not explicitly shown here. 

In linear theory, the velocity field will be proportional to the peculiar acceleration, which traces the spatial gradients of the gravitational potential. This is why the transverse velocity field is coupled to the underlying large-scale gravitational potential, so that the velocity can serve as a probe of gravitational effects in cosmology. We now give a general discussion of this phenomenon.

\subsection{Dipoles and transverse peculiar velocities}
\label{sec:dipole_prediction}

It is well known (e.g. section 8 of \citejap{Peebles1980}) that in linear theory the peculiar velocity is parallel to the peculiar gravitational field, which is in turn parallel to the integrated dipole of the density field:
\[
a\v{u} = {H(a)\fg(a)\over 4\pi}
\int {\delta(\v{r})\over (ar)^2}\, \v{\hat r}\, a^3d^3r.
\]
This equation is widely used to predict the peculiar velocities that should exist within regions of space mapped by galaxy surveys, estimating $\delta$ for the matter by using the galaxy number density fluctuation, which on large scales is $b\delta$, where $b$ is the linear bias parameter. Hence peculiar velocities can be predicted to within the unknown scaling factor $\fg/b$, and these predictions correlate very well with direct estimates of peculiar velocities within local volumes (e.g. \citejap{1988MNRAS.234..677L}; \citejap{DavisNusser2016}). In more distant surveys, the predictions can still be made, but here the interest is in the displacement field, $\v{D}$, where $\v{u}=d\v{D}/dt$. As discussed earlier, this is used in `reconstruction' analyses to make the BAO correlation signal more nearly linear \citep{2012MNRAS.427.2132P}; we exploit the data from such reconstructions here.

The interesting general phenomenon is that the peculiar velocity correlates not only with the dipole in the density field, but also with the dipole in the gravitational potential (and indeed with the dipole in any other field that has a wavelength-dependent linear relation to density). We will in practice use the density-inferred peculiar velocity as a `signpost' for these additional dipole effects, although in reality we are effectively correlating the density dipole with that in other fields. In principle all this can be done in 3D, and \cite{Ravoux2025} have recently written down the velocity and density correlations along the line of sight; but we will not do this, for two reasons. Firstly, redshift-space distortions of the apparent galaxy density field mean that the radial component of the peculiar velocity risks being mis-estimated relative to the transverse components. But also, data on relevant physical effects from the gravitational potential tend to come in projection -- especially in terms of foreground effects on the CMB. We therefore concentrate on dipoles that correlate with the transverse component of the peculiar velocity. To reveal these, we take maps of the quantity of interest and stack them about each galaxy in a given parent catalogue -- but rotated according to the direction of the transverse velocity at each stacking centre. In so doing, the magnitude of the velocity and its uncertain scaling with $\fg/b$ is irrelevant: we only care about the direction. This means that the cross-correlation between the velocity field and the gravitational potential is free from galaxy bias, and therefore partially free from the sample variance of the foreground large-scale structure, as we discuss in detail below.

We now give an overview of how such stacked fields should behave. In linear theory, the velocity will have a joint Gaussian distribution with density, potential etc. 
Density and velocity are uncorrelated at a given point, but they become correlated at non-zero separation, reasonably enough: the velocity field will tend to point from low density to high, purely as a consequence of continuity. For a point with position vector $\vec{r}$, the density fluctuation $\delta(\vec{r})$ is correlated with the component of velocity at the origin that is parallel to $\vec{r}$, but not with the perpendicular component.

For any two jointly Gaussian zero-mean quantities, $a$ \& $b$, the expected value of $b$ given $a$ is
\[
\langle b\mid a \rangle = {C_{ab}\over C_{aa}}\, a,
\]
requiring a knowledge of the covariance between $a$ \& $b$, $C_{ab}$, and the variance in $a$, $C_{aa}$. We now calculate this for the case of the correlation between the transverse velocity and the density in a 2D $xy$ slice perpendicular to the line of sight. Let the origin be where the velocity is measured, and consider a point offset by a distance $x$ in the $\hat x$ direction, where we wish to compute the density $\langle\delta(x)\rangle$ given the velocity $u_x$. If the density is a Fourier sum with components $\delta_k$, then
\[
u_x(0) = -iH\fg\sum_k \delta_k\, k_x/k^2,
\]
so that
\[
\eqalign{
C_{\delta u_x}\equiv\langle\delta^*(x) u_x(0)\rangle&=
-iH\fg \sum_k P(k)\, k_x/k^2\, \exp[ik_x x]
\cr
&=H \fg\sum_k P(k)\, k_x/k^2\, \sin[k_x x]
}
\label{eq:d-u-xcorr}
\]
and 
\[
C_{u_x u_x}\equiv\langle u_x(0)^2\rangle=
(H \fg)^2\sum_k P(k)\, k_x^2/k^4.
\]
The sums can be converted to $k$-space integrals, so that
\[
C_{\delta u_x}=
{H \fg\over (2\pi)^3}
\int\!\!\!\int k^{-2}P(k)\, k\mu\, \sin[k\mu x]\, 2\pi k^2\, dk\, d\mu
\]
and
\[
C_{u_x u_x}={(H \fg)^2\over (2\pi)^3}\int\!\!\!\int k^{-4}P(k)\, (k\mu)^2\, 2\pi k^2\, dk\, d\mu,
\]
where we use $k_x$ as a polar axis. Performing the $\mu$ integral gives 
\[
C_{\delta u_x}=H\fg \int k^{-1}\, \Delta^2(k)\, j_1(kx)  d\ln k
\]
and the 1D variance in the velocity at the origin is
\[
C_{u_x u_x}(0)=
(H \fg)^2
\int k^{-2}\, \Delta^2(k)/3\, d\ln k \equiv \sigma^2_{\japsc 1D}, 
\]
where $j_1(kx)=(kx)^{-2}[\sin (kx) - kx\,\cos (kx)]$ is the first-order spherical Bessel function.

\begin{figure*}
\begin{center}
	\includegraphics[width=0.47\textwidth]{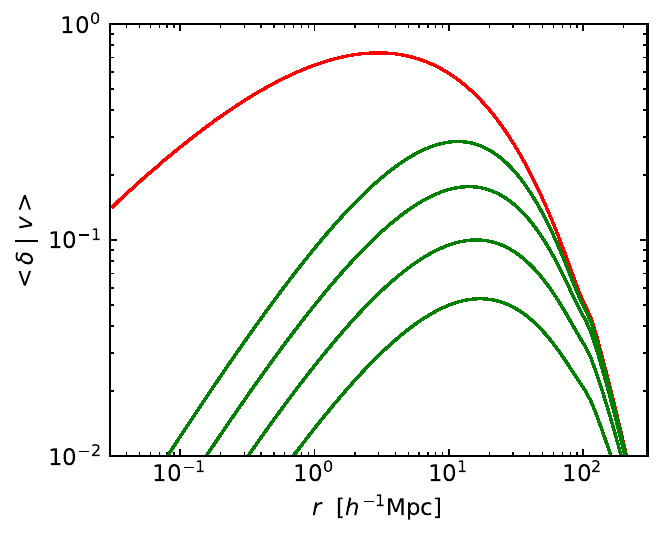}
    \includegraphics[width=0.47\textwidth]{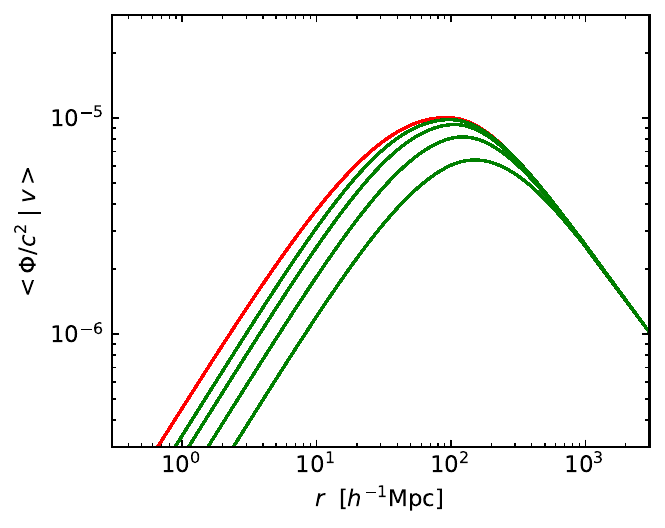}
    \caption{The dipole amplitude according to Eq.~(\ref{eq:dipole_amplitude}). This is calculated at $z=0$ according to linear theory, and scales in proportion to the linear fluctuation growth factor. The LH panel shows the case of density; the RH panel shows the case of gravitational potential, allowing for a flip in direction. Red lines show the idealised case of an infinitesimal slice, while green lines show the effect of nonzero extent along the line of sight. In the latter case, uniform slabs are considered, of thickness 50, 100, 200, $400\mpcoh$.
    }
    \label{fig:scalar_corr}
\end{center}
\end{figure*}

Similarly, the correlation between $\delta$ and $u_y$ is zero for separations in the $x$ direction.
If we then write $u_x = u_\perp \cos\theta$, where $\theta$ is the angle in the $xy$ plane between the transverse velocity and the direction of interest, the expected density at $x$ conditional on the transverse velocity at the origin then depends just on $u_x$ and thus shows a dipole:
\begin{equation}
    \langle \delta(\vec{r})\mid u_\perp \rangle 
    =\cos\theta \times u_\perp 
     C_{u_x \delta}/\sigma^2_{\japsc 1D}.
\end{equation}
We now need to average this again, over velocity. For 2D fields in the plane of the sky, where we ignore radial components, we have
$\langle |u_\perp| \rangle = (\pi/2)^{1/2}\sigma_{\japsc 1D}$. But this is not a very practical form of averaging: in estimating the dipole in real data we should downweight points with low $u_\perp$, since they will only add noise. We choose to perform an average weighted according to velocity, which involves the mean of $u_\perp^2$, $2\sigma_{\japsc 1D}^2$, with the mean of $|u_\perp|$ in the denominator as the mean weight.
Thus the mean density dipole is
\begin{equation}
\langle \delta(\vec{r})\mid \vec{u} \rangle_{u {\rm weight}} =
\cos\theta \times (8/\pi)^{1/2} C_{u_x\delta}(x)/\sigma_{\japsc 1D}.
\label{eq:dipole_amplitude}
\end{equation}
A plot of this dipole amplitude as a function of separation is given in Fig.~\ref{fig:scalar_corr}.

From this derivation, it should be clear that there is a corresponding dipole
in any quantity that has an isotropic linear relation to density in Fourier space.
If $X_k(\v{k})= g(k)\delta_k(\v{k})$, then the above expression for the dipole still
holds, except that $\Delta^2(k)$ needs to be replaced by $g(k)\Delta^2(k)$ in the expression for $C_{u_x\delta}$. The prime example of this is the gravitational potential, where \[
\Phi_k = g(k)\delta_k =-(3/2)\Omega_m H_0^2 a^{-1} k^{-2}\, \delta_k. 
\]
We see that the sign of the potential dipole will be opposite to that of density. reasonably enough: velocity points towards a positive overdensity, which is a potential dip.  
The case of the potential dipole amplitude is also shown in Fig.~\ref{fig:scalar_corr}, where we disregard this minus sign.

\subsection{A signal independent of galaxy bias}

The above analysis shows that there will be a dipole in density and in potential if we stack maps of such fields about the direction of the peculiar velocity. This result is derived assuming that the points at which the velocity is measured are randomly selected. But it is convenient in practice to use the positions of foreground galaxies as stacking centres. This introduces a weighting factor of $1+\delta_g$, and this factor means that there will then also be a monopole arising from the cross-correlation between $\delta_g$ and the field being stacked. This monopole is not to be confused with the term `monopole' as used in the context of the CMB, where it means just the all-sky average temperature. Here, the term monopole refers to the conventional cross/auto-correlation function, characterising the amplitude of fluctuations as a function of (angular) scale. In terms of the rotated and stacked maps of CMB temperature etc., the monopole represents the average over directions within an annulus of given radius; it is the natural independent counterpart to the dipole signal, which measures the anisotropy within the same annulus.

It is interesting to contrast the parameter sensitivities of these multipoles, particularly in terms of galaxy bias, $b$. If we estimate transverse peculiar velocities from a galaxy sample, and stack the galaxy number density field itself using galaxies as stacking centres, the monopole will scale as $b^2\sigma_8^2$,
or as $b\sigma_8^2$ when stacking (or correlating with) pure dark-matter quantities such as potential or lensing convergence. But the dipole has a weaker (or no) dependence on bias, as may be seen from Eq.~(\ref{eq:dipole_amplitude}). There, $\sigma_{\rm 1D}$ is the dispersion in velocity, which scales as $Hf\sigma_8$, whereas $C_{u\delta}$ scales as $bHf\sigma_8^2$ if stacking galaxy density, so that the dipole scales as $b\sigma_8$. The density-velocity coupling prefactor $Hf$ cancels out since we are only using the direction of the velocity in the stacking. 
But if we are stacking maps related to potential or lensing convergence, then there is no factor of $b$, and we have the powerful result that the stacked dipole depends purely on the properties of matter, independent of the bias of the galaxies used as stacking centres -- which is not true for the monopole.

We have tested this explicitly with N-body simulations by volume weighting the data: sampling the $\bf u$ field uniformly, so that the mean value of $\bf u$ is zero. We recover the same dipole in density or potential as in the case where we sample $\bf u$ only at the locations of haloes when the weights associated with the number of haloes are accounted for. 
But the behaviour is completely different for the monopole, which vanishes when we average around randomly selected points.
In the following comparisons between simulations and linear theory, we will show dipoles produced using equal-volume sampling of the velocity field; but we will show multipoles estimated from observational data in which the extra $(1+\delta_g)$ weighting is applied.

Lastly, it is also self-evident that the dipole is independent of the monopole. Analysing the dipole signal should therefore yield additional cosmological information, beyond what we learn from the conventional cross-correlation functions. 

\subsection{Dipoles and monopoles in projection}

\label{sec:slab_projection}

The above case, with the transverse velocity and the scalar quantity of interest measured in a sheet, is doubly unrealistic. The transverse velocity is estimated within a 3D galaxy survey, so any centres chosen for stacking will inevitably exist over some range of distances set by the survey redshift distribution. Furthermore, the quantity that has a dipole relation to the transverse velocity may have a kernel that is distributed in depth. The simplest case would be stacking the number density of galaxies on the sky aligned with the peculiar velocity, in which case the effect of the survey redshift distribution will enter twice.

As we discuss below, we will be interested in stacking 2D fields that are correlated with the projected distribution of galaxies. This stacking can either be performed with rotation in order to align the transverse velocities, in which case, we expect to see a dipole signal; or we can stack without rotation -- in which case we obtain the monopole signal, which represents the mean effect as a function of radius: this is just the usual angular correlation function. The fields of interest are the projected density and potential, both of which are in the form of a radial integral involving some kernel:
\[
\bar\delta(x,y) = \int \delta(x,y,r)\, K(r)\, dr,
\]
and similarly for $\bar\Phi$, where here and below $r$ denotes comoving radius; the $x$ \& $y$ coordinates are transverse to the line of sight.

This averaged density at $y=0$ is written in Fourier space as
\[
\bar\delta(x) = \sum_k \delta_k\, \exp[-ik_x x]\, \tilde K_1(k_z).
\]
We now wish to correlate this with the velocity field, which may be at a location offset by an amount $z$ radially from the centre of the slab (taken to be $z=0$).
In that case, the correlation is
\[
C_{u_x\bar\delta}=
Hf \sum_k P(k) \, {k_x\over k^2}\, \sin[k_x x] \,\tilde K_1(k_z)\, \cos[k_z z],
\]
where for simplicity we assume that the kernel is symmetric, so that its transform can be taken to be real. If we now average this correlation for positions of centres distributed radially with kernel $K_2$, then integration over $z$ converts the $\cos[k_z z]$ term into $\tilde K_2(k_z)$. This expression for $C_{u_x\bar\delta}$ is of the identical form to the one that we would have obtained if we had first averaged $\delta$ and $u_x$ within separate slabs and then correlated the two resulting 2D fields. But these two operations are subtly different, since LOS averaging will cause a slight reduction in the amplitude of transverse velocities. The equation for the dipole, Eq.~(\ref{eq:dipole_amplitude}) shows that  the amplitude is proportional to $C_{u_x\delta}/\sigma_{\rm 1D}$; thus the reduction in $\sigma_{\rm 1D}$ means that we will obtain a different dipole amplitude if we average the transverse velocities prior to stacking. We do not follow this route in the present paper, but it may be worth further investigation.

To evaluate the sum in the expression for the correlation as an integral, we choose $k_z$ as the polar axis in $k$-space, so that $k_z=k\mu$. Performing the azimuthal integral then yields
\[
\label{eq:xi_u_aniso}
\eqalign{
C_{\delta u_x}(x) &= H\fg 
\int k^{-1}\Delta^2(k)\, d\ln k \; \times \cr
&{1\over 2}\int_{-1}^1 d\mu\, \tilde K_1(k\mu) \tilde K_2(k\mu)\, \sqrt{1-\mu^2}\, J_1[\sqrt{1-\mu^2}\, kx],
}
\]
where $J_1$ is a normal Bessel function.
For the monopole signal, we require the density autocorrelation $C_{\bar\delta_1\bar\delta_2}$, which is obtained by similar reasoning:
\[
\eqalign{
C_{\bar\delta_1\bar\delta_2}(x) &=
\int_0^\infty \Delta^2(k)\; d\ln k\;\times
\cr
&{1\over 2}\int_{-1}^1\tilde K_1(k\mu) \tilde K_2(k\mu)
\,J_0(\sqrt{1-\mu^2}\,k\,x)\; d\mu.
}
\]
If we are interested instead in stacking potential fields, these expressions are simply modified by inserting a single power of the factor that relates $\Phi_k$ to $\delta_k$.

As a simple example of the application of these formulae, suppose that both velocity and density are averaged in a slab of thickness $L$: then 
\[
\tilde K_1 = \tilde K_2 = {\sin[k\mu L/2]\over k\mu L/2}.
\]
Fig.~\ref{fig:scalar_corr} illustrates the effect of finite thickness using this formula.

\subsection{Limber approximation}

\label{sec:limber_prediction}

Now, for comparison to observations, we will be concerned with the correlations as a function of angular separation, rather than spatial offset. There are two ways of implementing this, both of which are approximations that should be valid at small angular separations. The first is to argue that the kernel functions pick out some preferred distance, $R$, which we can take to be the mean distance for the galaxy sample used in the correlation measurements. Then the slab expressions should apply with a transverse separation that can be expressed in angular terms:
\[
x \to R\theta;\quad
R = \int r\, K_{\rm gal}\, dr,
\]
where $\theta$ now denotes angular separation, not the azimuthal angle with respect to the velocity direction.

Alternatively, we can note that the kernels are still broad by comparison with the typical scales of non-zero spatial correlations. In that case we have the \cite{Limber1953} approximation, in which radial integrals are taken over an infinite range:
\[
C_{\delta_1\delta_2}=\int\!\!\!\int \xi({\bf r}_1-{\bf r}_2)\, K_1(r_1) K_2(r_2) \,dr_1 dr_2,
\]
where the kernels are written in terms of the redshift distributions as $K(r)=p(Z)\, (dZ/dr)$ (making redshift a capital letter to avoid confusion with the radial direction). For thick slices, the signal is dominated by $r_1\simeq r_2$, so that $\int K_1(r) K_2(r) \,dr_1 dr_2$ is replaced by $\int K_1(r) K_2(r)\, dr\, dz$, where $z=r_1-r_2$ .
For a small-angle approximation, we write the spatial correlation as $\xi(z^2+r^2\theta^2]^{1/2})$. Writing the correlation function in Fourier space allows the $z$ integral to be performed, yielding
\[
C_{\delta_1\delta_2}(\theta)=
\pi\int K_1(r)K_2(r)\; dr\int \Delta^2(k)\, 
J_0(kr\theta)\;{dk\over k^2}
\label{eq:limber0}
\]
for the monopole. For the dipole, we proceed similarly, but need to allow for the correlations being anisotropic, as in Eq.~(\ref{eq:xi_u_aniso}). If we write the sum over modes as an integral with a $k$-space volume element $k^2 dk\, d\phi \, d\mu/2$, with $k_z=\mu k$ and $k_x=k\sqrt{1-\mu^2} \cos\phi$, then the radial integral gives a $\delta_{\rm D}$-function: $\int \exp[ik\mu z]\, dz = (2\pi/k)\delta_{\rm D}(\mu)$. Performing the integral over $\phi$ yields a final expression of similar form to the one for the monopole:
\[
C_{\delta u}(\theta)=
\pi H \fg\int K_\delta(r)K_u(r)\; dr \int \Delta^2(k)\, 
J_1(kr\theta)\;{dk\over k^3}\, .
\label{eq:limber1}
\]
To correlate transverse velocity with other fields, such as gravitational potential, one would use the same expression, but introducing appropriate $k$-space factors to convert from density to the field of interest.

To go beyond the small-angle approximation, we should at a minimum replace $\theta$ by $2\sin(\theta/2)$. But these two functions of $\theta$ differ by only 1\% even at $\theta=30^\circ$. In more detail, we have compared the slab dipoles shown in Fig.~\ref{fig:scalar_corr} with the same quantities computed in the Limber approximation. For a slice thickness of relevance for the data used here (approximately $400\mpcoh$), the two expressions agree to within a few \% up to separations of $10^\circ$, which we therefore use as the upper limit when comparing data to the Limber prediction. 

We should note that all these formulae include an implicit time dependence of the matter power spectrum: $\Delta^2(k)\to\Delta^2(k,r)$. In the limit of linear evolution, this amounts to multiplying the kernel function by the linear growth law as a function of distance (and also to bringing the distance-dependent $H\fg$ inside the radial integral).

\section{Dipoles in the ISW effect and gravitational lensing}
\label{sec:ISW_Lensing_Dipoles}

We now look in more detail at the relation between the ISW effect and gravitational lensing from the perspective of transverse velocities. We will demonstrate that transverse velocities naturally connect these gravitational effects, providing alternative observational probes of the largest-scale perturbation modes. 

\begin{figure*}
\begin{center}
\scalebox{1.0}{
	\includegraphics[width=0.99\textwidth]{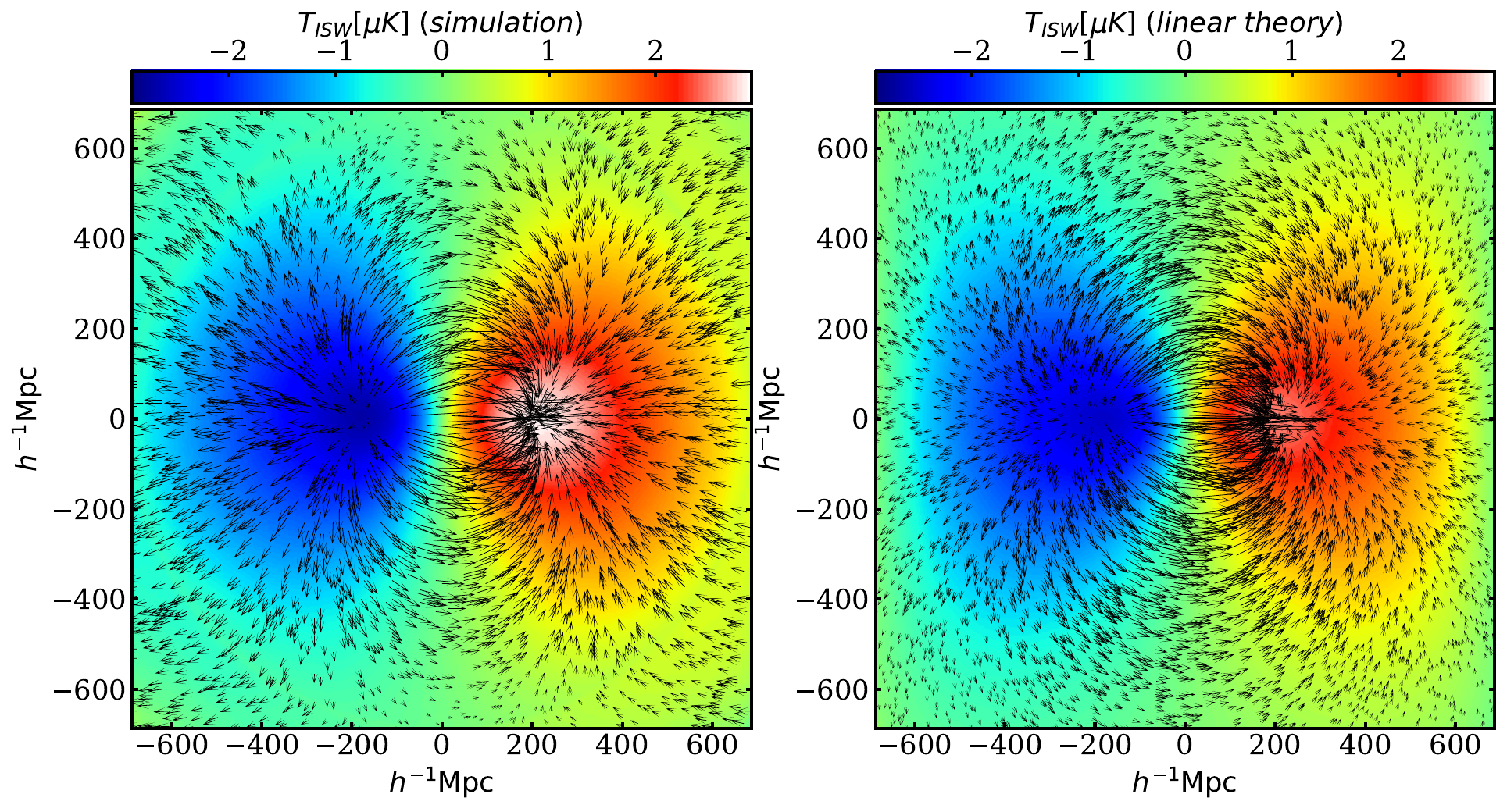}}
    \caption{The ISW temperature fluctuations induced by the large-scale velocity field shown in Fig.~\ref{v-correlation-2D}, calculated using Eq.~(\ref{eq:v-ISW}). The signal is integrated over a slab of LOS depth $400\mpcoh$. Left: measurement from N-body simulations, as as discussed in Appendix \ref{appA}; right: linear theory. For a closer comparison with BOSS data, this figure shows results at $z=0.55$, rather than $z=0$.
    }
    \label{Tisw-2D}
\end{center}
\end{figure*}

\begin{figure*}
\begin{center}
\scalebox{1.0}{
	\includegraphics[width=0.99\textwidth]{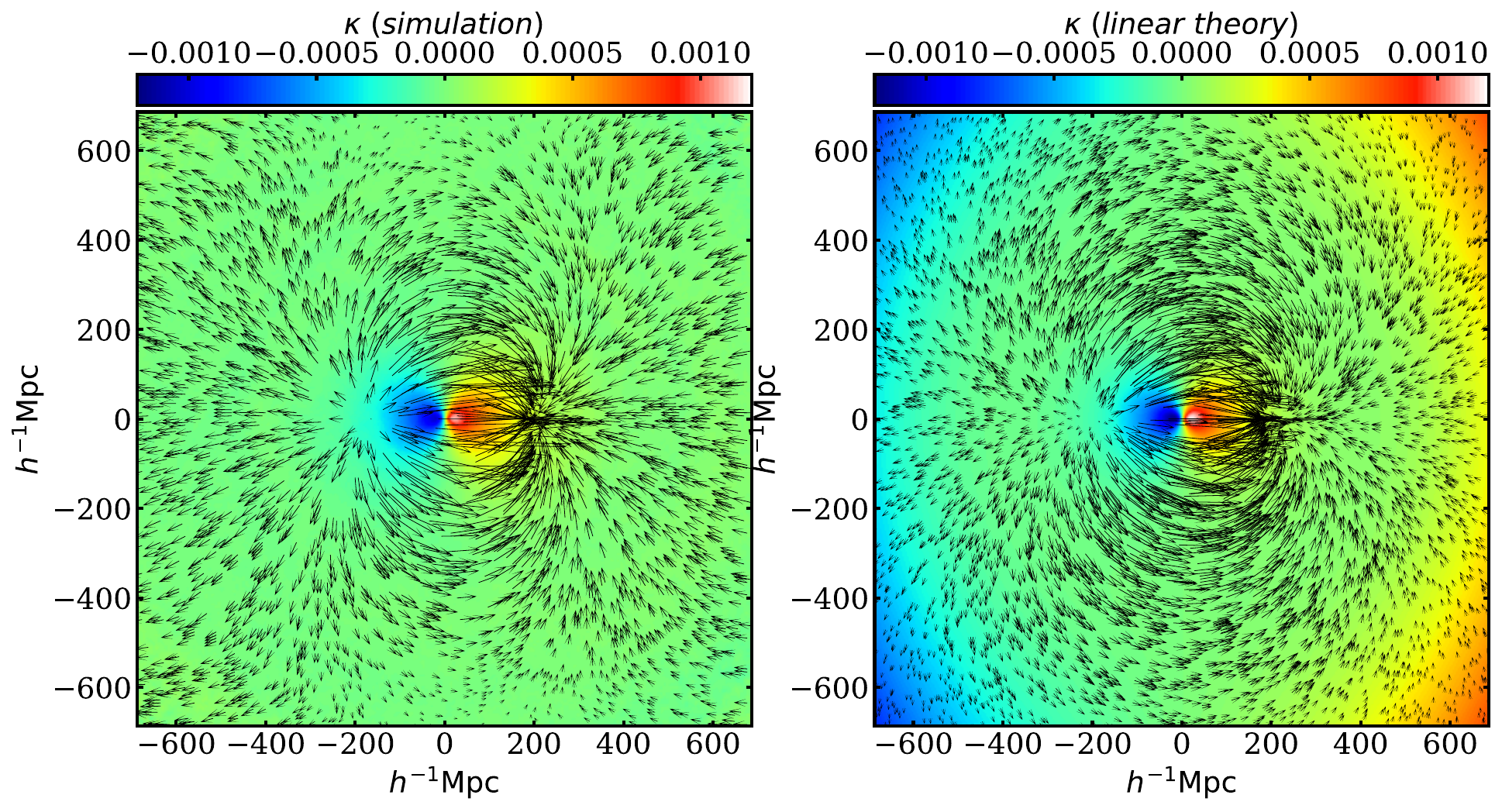}}
    \caption{The expected CMB lensing convergence map induced by the large-scale velocity field shown in Fig.~\ref{v-correlation-2D}, but now computed at $z=0.55$. The signal is integrated over a slab of LOS depth $400\mpcoh$. Left: measurement from N-body simulations, as as discussed in Appendix \ref{appA}; right: linear theory calculated using Eq.~(\ref{eq:dipole_amplitude}), but replacing $\sqrt{8/\pi}$ by $\sqrt{\pi/2}$ for the case of equal weighting.}  
    \label{kappa-2D}
\end{center}
\end{figure*}

\subsection{Velocity-ISW correlations}

\subsubsection{ISW in configuration space}
The ISW effect arises from the time derivative of the gravitational potential. Using comoving coordinates $\bf x$, the potential is
\[
\Phi (\vec x)= -G \int d^3 x'\, a^3\, {\bar\rho\, \delta({\bf x'})
\over a\, |{\bf x-x'}|} =
{-G\bar\rho_0 \over a} \int d^3 x'\, {\delta({\bf x'})
\over |{\bf x-x'}|},
\]
where $G$ is the gravitational constant, $a$ is the cosmic scale factor, $\bar \rho$ is the mean density at $a$ and $\bar \rho_0$ is its value at $a=1$.
Hence the time derivative is
\[
\label{eq:LinearISW_Real}
\dot\Phi (\vec x)= -{\dot a\over a}\, \Phi(\vec x)  - {G\bar\rho_0 \over a} \int d^3 x'\, {\dot\delta({\bf x'})
\over |{\bf x-x'}|} .
\]

An alternative form of this expression can be obtained by using the divergence theorem applied to the continuity equation $\dot\delta = -(1/a){\bf\nabla\cdot} (1+\delta) {\bf v}$, as suggested by \citet{Rubino-Martin2004}. Note once again that spatial derivatives are comoving, but the peculiar velocity is proper. 
The second integral is of the form 
$\int d^3x\, f{\bf \nabla\cdot F}$, which is $-\int d^3x\, {\bf F\cdot \nabla}f$ if we use the divergence theorem for $f{\bf F}$ and argue that the boundary term at infinity is negligible. This gives a two-part expression for the time derivative of the potential:
\begin{equation}
\label{eq:general_phidot}
\dot\Phi (\vec x)= -{\dot a\over a}\, \Phi(\vec x) - {G\bar\rho_0 \over a^2} \int d^3 x'\, {(1+\delta)\,\bf v \cdot (x-x')\over |{\bf x-x'}|^3}.
\end{equation}
This general expression for the effect of evolving potentials allows a calculation of both linear ISW and nonlinear Rees-Sciama effects on the CMB (see e.g. \citejap{Cooray2002}), which are given without approximation by 
\[
\eqalign{
 &{\Delta T(\vec x) \over T_{\japsc CMB}} =-\frac{2}{c^3}\int a\, dr \frac{\dot a}{a}\Phi (\vec x) \cr
 \,&-  {2 G\bar\rho_0 \over c^3} \int a\, dr \left[\int d^3x' \, {(1+\delta){\bf v \cdot (x - x')}\over a^2 |{\bf x - x' }|^3}\right].
 }
\]
Here it is understood that $\Phi$, $\delta$ and $\vec v$ are all exact quantities, including effects of nonlinear evolution.
Performing the radial integral yields
\begin{equation}
\label{eq:ISW}
\eqalign{
 {\Delta T (\vec x) \over T_{\japsc CMB}}&= -\frac{2}{c^3} \int \dot a\, dr \, \Phi(\vec x) \cr
&- {4 G\bar\rho_0 \over c^3} \int d^2x'_\perp \, {\bf p(x_\perp) \cdot (x_\perp - x'_\perp)\over |{\bf x_\perp - x'_\perp }|^2},
}
\end{equation}
where the projected momentum field is
\begin{equation}
\label{eq:momentum}
{ \vec p(\bf x_\perp)} = \int dz \, [1+\delta(\vec x)] {\bf v(\vec x)} /a.
\end{equation}
The factor 2 arises because we can write
\begin{equation}
|{\bf x - x' }|^3
= \left(
|{\bf x_\perp - x'_\perp }|^2 + Z^2
\right)^{3/2},
\end{equation}
where $Z\equiv z-z'$. The radial integral is therefore of the form
\begin{equation}
\int_{-\infty}^\infty 
{dZ\over (X^2 + Z^2)^{3/2}},
\end{equation}
which is $2/X^2$.

This general expression is however more than we need for many applications, where fluctuations are close to the linear regime. The potential obeys the comoving Poisson equation:
\begin{equation}\label{eq:Poisson_Real}
\nabla^2\Phi({\vec x,t})=4\pi G
\bar{\rho}(t) [a(t)]^2\delta({\vec x, t}),
\end{equation}
where $\bar\rho(t) = a^{-3}\bar\rho_0$; in the linear regime we have
$\delta({\vec x}, t)\propto D(t) \delta({\vec x}, t_0)$,
where $D(a)$ is the linear density growth factor. Hence, in the linear regime, we have
$\Phi(a)\propto D(a)/a$, so that
\begin{equation}
\label{eq:ISWLin}
\dot\Phi = H(a) (\fg-1)\Phi.
\end{equation}
Thus the second term in Eq.~(\ref{eq:general_phidot}) is approximately $H\fg\Phi$,  which is $\dot\Phi\, \fg/(\fg-1)$. Hence in situations where linear theory applies, we can write the ISW effect entirely in terms of the second term in Eq.~ (\ref{eq:ISW}). This has the practical advantage that we only need the velocity field and are spared the need to estimate the potential explicitly. In this linear regime, the expression for
$\bf p(\vec x_\perp)$
simplifies
and the ISW effect is determined purely by the transverse velocity field:
\[
\eqalign{
\frac{\Delta T(\vec x_\perp)}{T_{\japsc CMB}} &= {\textstyle\left({1\over \fg}-1\right)}\,\frac{4G\bar\rho_0}{ac^3} \!\int\! dr\!\int\! dx_\perp'^2 \vec v({\vec x_\perp')} {\bf\cdot} \frac{{\bf x_\perp}-{\bf x_\perp'}}{|{\bf x_\perp}-{\bf x_\perp'}|^2} \cr
&\kern-3.5em = {\textstyle\left({1\over \fg}-1\right)}\, \frac{3H_0^2\Omega_m}{2\pi a c^3} \!\int\! dr \!\int\! dx_\perp'^2 \vec v({\vec x_\perp')} {\bf\cdot} \frac{{\bf x_\perp}-{\bf x_\perp'}}{|{\bf x_\perp}-{\bf x_\perp'}|^2}\, .
}
\label{eq:v-ISW}
\]
Thus the expression given by \citet{Rubino-Martin2004} needs to be supplemented by a large-scale term corresponding to the expansion-driven decay of the potential (the dominant effect at late times in a $\Lambda$-dominated model, once growth of $\delta$ slows down and 
$\fg\rightarrow 0$).

The ISW temperature dipole about a given point is thus aligned with the transverse velocity at that point. In section~\ref{sec:vcorrelations}, we derived Eq.~(\ref{eq:v2D}), which shows how the transverse velocity field depends on position relative to the transverse velocity at a given point, which was illustrated in Fig.~\ref{v-correlation-2D}. 
We insert this linear velocity field into Eq.~(\ref{eq:v-ISW}) in order to calculate the ISW dipole in a slab centred at $z=0.55$, which is shown in the left-hand panel of Fig.~\ref{Tisw-2D}. We can see that the coherence scale of the dipole extends to several hundred Mpc. The divergence and convergence flows of the velocity field corresponds very well with the cold and hot spot of the temperature dipole, with the amplitudes of fluctuations up to a few$\,\mu$K. The dipole peaks at around $200\mpcoh$ as shown in the bottom panel of the figure. The prediction agrees very well with measurements from N-body simulations shown in the left-hand panel.

The temperature dipole of the ISW effect and its association with the transverse velocity field offers a novel means for detecting the effect. If we average the ISW signal azimuthally about a given point, we would obtain the standard ISW monopole signal, which has been the subject of much study. But the dipole is statistically independent of the monopole and provides additional independent cosmological information. Indeed, as we will show below, the dipole provides a substantially more sensitive probe of the ISW effect.

Beyond mere $S/N$, the ISW dipole has one further advantage over the monopole. It is obvious from Eqs.~(\ref{eq:psi}) \& (\ref{eq:v-ISW}) that the ISW dipole depends directly on the matter density parameter $\Omega_{\rm m}$, and all other cosmological parameters that influence the velocity and density power spectra. However, because only the {\it direction\/} of the estimated peculiar velocity is used, the ISW dipole is independent of galaxy bias, and it can therefore provide a direct probe of the amplitude of the matter or velocity power spectrum. In addition, due to its direct connection to the velocity field, the dipole receives contributions mainly from the very large-scale perturbation modes. This makes it a useful probe of non-Gaussianity and effects of general relativity that are typically prominent only for the very low-$k$ modes, again free from the effects of galaxy bias.

\begin{figure*}
\begin{center}

\scalebox{1.0}{
	\includegraphics[width=0.99\textwidth]{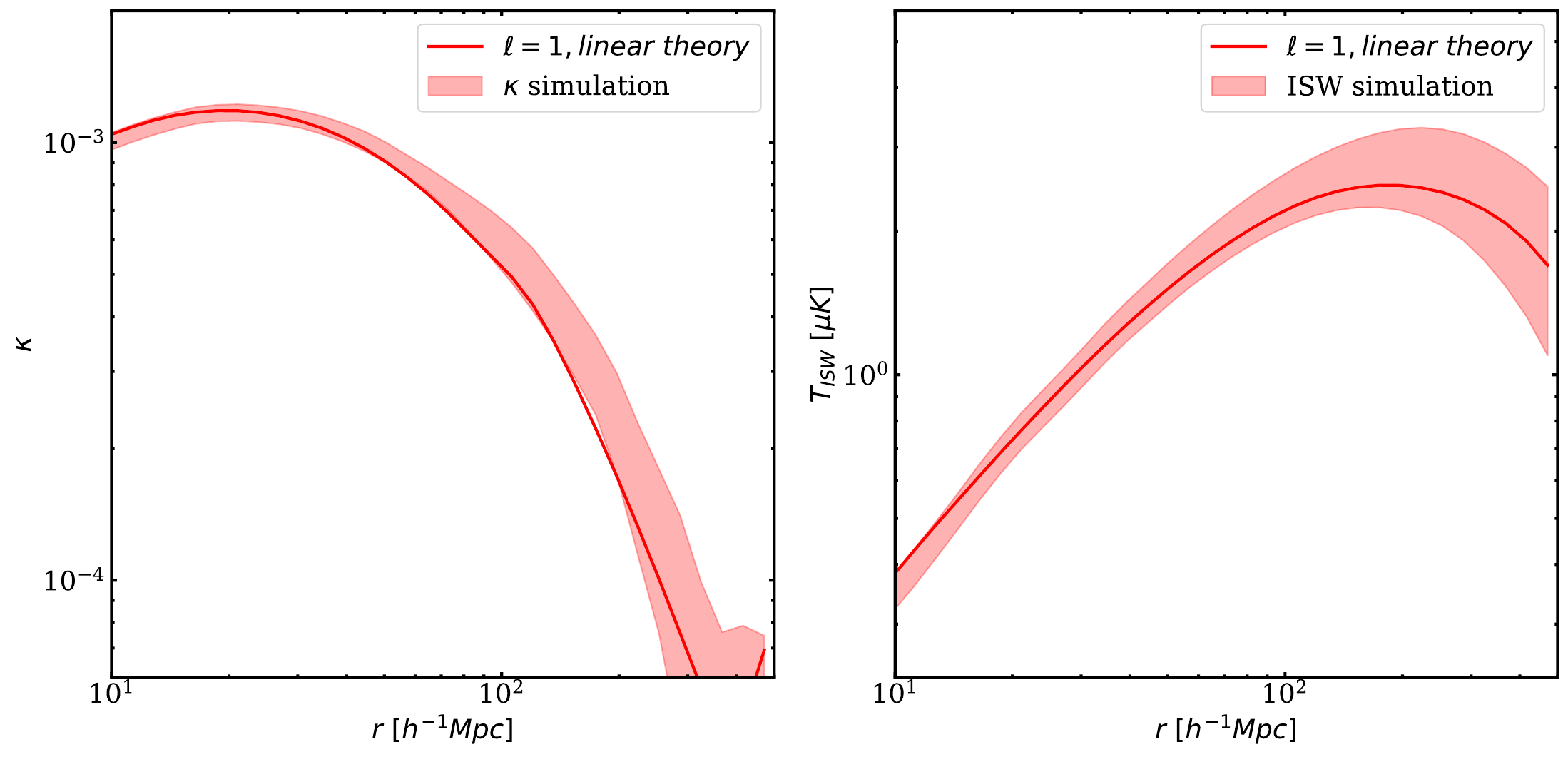}}
    \caption{Left: Comparing dipoles of the lensing convergence between linear theory and the N-body simulations shown in Fig.~\ref{kappa-2D}. Right: comparing dipoles of the ISW temperature fluctuations shown in Fig.~\ref{Tisw-2D}. The dipoles are computed at $z=0.55$ from the simulated maps using Eq.~(\ref{eq:decomposition}).}
    \label{Dipoles_kappa-ISW}
\end{center}
\end{figure*}

\subsection{Velocity-lensing correlations}

\subsubsection{Comparison with the moving lens effect}

\label{sec:moving_lens}

The ISW effect as discussed above is largely a linear and large-scale effect.
The velocity vector at the stacking centre serves to locate large-scale gravitational potential wells and potential hills, which will decay once the universe becomes $\Lambda$-dominated: it is this time variation  that dominates the ISW dipole, and it exists only when $\fg \neq 1$. 
Conversely, a local potential induced by mass at the stacking centre will generate both a nonlinear contribution to the ISW monopole and also an additional nonlinear dipole: the moving gravitational lens effect first pointed out by \cite{Birkinshaw1983}.
 
We can look at the case of a moving gravitational potential well from the viewpoint of the ISW effect. Ahead of the well, where $\vec v {\bf\cdot} ({\bf x_\perp}-{\bf x_\perp'}) >0$, the potential is becoming deeper with time. This will cause the traversing CMB photons to lose some energy, and thus induce a CMB cold spot. Behind the well, however, the potential is becoming shallower with time and so the effect is opposite \citep{Cai2010}. The CMB thus displays a dipole that is anti-aligned with the transverse motion: opposite in sign to the linear ISW effect discussed above. In that case, the velocity points towards decaying potential wells, for which $\dot\Phi>0$. Note that the typical scale of the moving-lens dipole is much smaller, $\sim$\,$10\mpcoh$, compared to the $\sim$\,$100\mpcoh$ ISW dipole. Its amplitude is also expected to be smaller, typically at the level of $\mu K$ or less \citep{Rubino-Martin2004, Cai2010, Yasini2019, Hotinli2021, Beheshti2024}.

The result for the moving well was derived in a completely different way by \cite{Birkinshaw1983}, who considered that photons deflected by the moving gravitational lens pick up a Doppler shift from a component of its transverse velocity. Their result for the amplitude of the resulting temperature dipole was
\begin{equation}
{\Delta T\over T} = \gamma (v_\perp/c)\alpha,
\end{equation}
where $\gamma$ is the Lorentz factor and $\alpha$ is the lensing bend angle. Here we have corrected what seems to be an error of a factor $-2$ in converting between fractional shifts in frequency and temperature; we believe that these two shifts should be identical.
As we now show, this expression arises naturally as an aspect of the ISW effect in terms of $\dot\Phi$, so the moving-lens dipole is not a distinct phenomenon. From the standard theory of weak gravitational lensing (e.g. \citejap{bartelmann2001}), the vector bend angle is
\begin{equation}
\vec{\alpha} = \vec{\nabla}_\perp\, 
{2\over c^2}\int \Phi\, dr
\end{equation}
(where all lengths are comoving). Now, $\vec{v}_\perp {\bf\cdot} \vec{\nabla} \Phi = a\dot\Phi$; the power of $a$ is absorbed via $a\, dr = c\, dt$, so indeed the moving-lens fractional temperature dipole is the standard ISW expression: $2/c^2\int \dot\Phi\, dt$ (for non-relativistic velocities with $\gamma\simeq 1$).

In summary, from the perspective of $\dot\Phi$, the moving well is an example of the post-linear ISW effects generally named after \cite{Rees1968}.
Its amplitude is directly proportional to the mass of the moving object at the centre, and its coherence scale is much smaller than that of the linear ISW effect \citep{Cai2010, Yasini2019,Hotinli2021,hotinli2023, hotinli2024}. But it is clearly distinct from the large-scale ISW dipole, even though both are aspects of time-varying potentials. The small-scale dipole arises from a moving well, whose motion points along the gradient in potential between a large-scale potential hill and a potential well. But these large-scale gravitational wells are not in motion, and change with time because the stretching effect of dark energy causes the potentials to decay. So we end with a small-scale moving-lens dipole embedded within a large-scale ISW dipole that has the opposite sign of temperature.

\subsubsection{Velocity-CMB lensing correlations}

Returning to the linear regime, there is a general connection between ISW temperature perturbations and the signal of weak lensing. This is defined using a lensing potential:
\begin{equation}
{r_{\rm S}- r_{\rm L}\over r_{\rm S} }\vec{\alpha} = \vec{\nabla}_\theta \psi_{\rm L};
\quad
\psi_{\rm L}\equiv {2\over c^2} \int {r_{\rm S}- r_{\rm L}\over r_{\rm S}\, r_{\rm L} }
\Phi\, dr_{\rm L},
\end{equation}
where $r_{\rm L}$ and $r_{\rm S}$ are the comoving distances to the lens plane and the source respectively. 
\cite{Hang2021b} showed that in the linear regime and for perturbations in a thin shell, the ISW temperature fluctuation is proportional to the lensing potential via 
\begin{equation}
\label{eq:ISW-lensing}
\frac{\Delta T(\vec \theta)}{T_{\japsc CMB}}=aH(a)[1-f(a)]\frac{r_{\rm S}\,r_{\rm L} }{c(r_{\rm S}- r_{\rm L})}\Psi_{\rm L}(\vec \theta),
\end{equation}
This, together with Eq.~(\ref{eq:v-ISW}) connects the velocity field with the lensing potential. Thus, there will be a dipole in the lensing potential, aligned with the peculiar velocity, just as there will be a dipole in the ISW signal. 

Alternatively, gravitational lensing may be specified in terms of the convergence, $\kappa$, which gives the RHS of the lensing Poisson equation: $\nabla^2_\theta \psi_{\rm L} = 2\kappa$. This convergence field is a suitably weighted line integral of the density field:
\begin{equation}
    \kappa = {3H_0^2\Omega_m\over 2c^2}
    \int_0^{r_{\japsc CMB}}\delta\, {r(r_{\japsc CMB}-r)\over a\, r_{\japsc CMB}} dr,
\end{equation}
where $r_{\japsc CMB}$ is the comoving distance to last scattering at $z\simeq1100$. Thus there will also be a dipole in the $\kappa$ map -- and in any fields of biased tracers of the projected matter density such as galaxies -- aligned with the transverse peculiar velocity. A prediction of the $\kappa$ stack in a slab at $z=0.55$, made on the same basis as the ISW prediction of Fig.~\ref{Tisw-2D}, is shown in Fig.~\ref{kappa-2D}. We can see that the dipole of the CMB lensing convergence is expected to be of the order of $10^{-4}$ to $ 10^{-3}$ for this redshift range, and is much more localised. The signal peaks at around 20$~h^{-1}$Mpc separation, and then decreases with scale.

As with the ISW signal, the dipole from the cross-correlation between transverse velocity direction and lensing is independent of galaxy bias, but its monopole (the conventional galaxy-lensing cross-correlation) depends linearly on galaxy bias. The lensing dipole alone can thus be used to constrain cosmology free from galaxy bias, while the combination of the lensing monopole and dipole can be used to constrain bias.

\subsection{Summary}

We have discussed the theoretical and statistical reasons why physical effects related to the gravitational field should show a dipole pattern in alignment with the peculiar velocity at a point. In particular, we have shown that both the ISW temperature perturbations of the CMB and gravitational lensing (quantified via  either the lensing potential or the convergence) should display a dipole that correlates with the transverse velocity field. Both these effects will also show a non-zero monopole if averaged about the locations of galaxies. But in that case, the monopoles will have an amplitude that scales with the bias parameter of the galaxy sample, whereas the dipoles are independent of bias. The dipole signal therefore forms an attractively novel observational probe, which is statistically independent of the monopole, as well as being more robust. We have shown in Section  \ref{sec:dipole_prediction} how to predict the ISW and lensing multipole amplitudes, and in particular given convenient expressions for them in Section \ref{sec:limber_prediction} using the Limber approximation. We now attempt to measure these effects in real observational data.

\begin{figure*}
\begin{center}
\scalebox{1.0}{
	\includegraphics[height=0.5\textwidth]
    {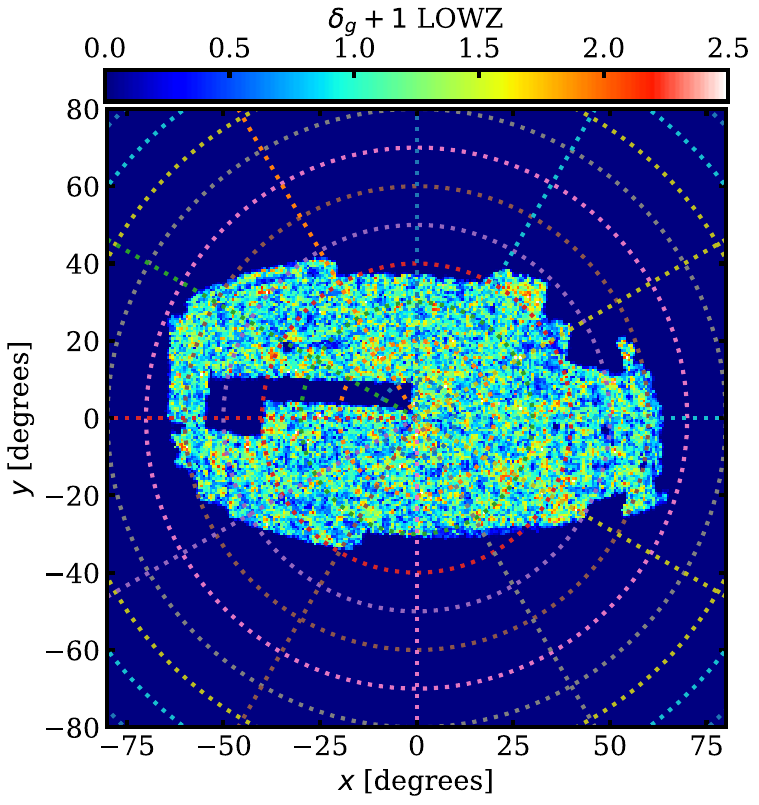}}
 \scalebox{1.0}{
 \includegraphics[height=0.5\textwidth]
    {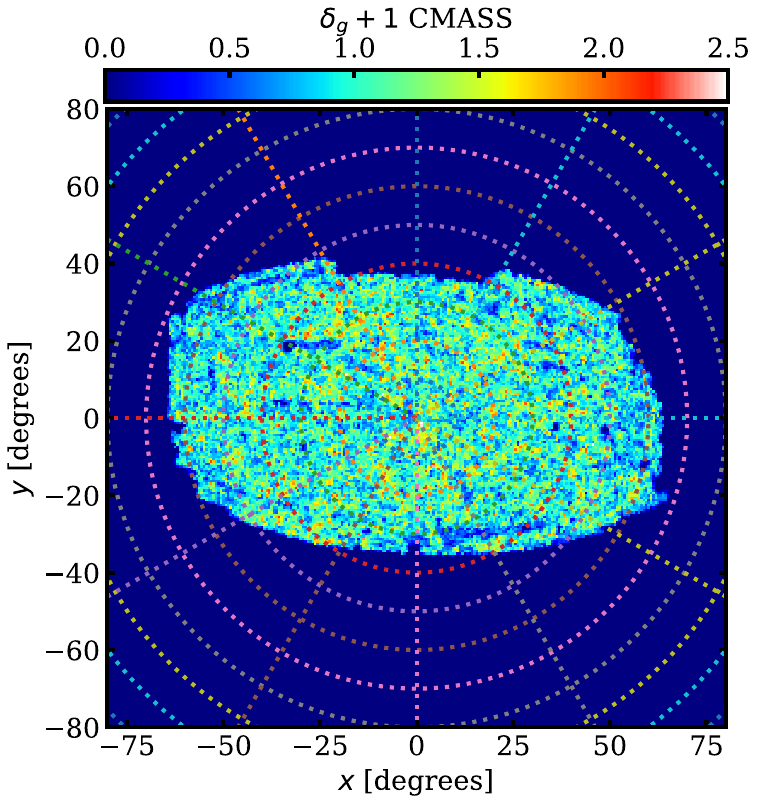}} 
    \caption{Azimuthal-equidistant projections of the galaxy number counts per $\mbox{\sc nside}=64$ {\sc Healpix} pixel \citep{Gorski2005} for the BOSS CMASS sample (left) and LOWZ sample (right). The maps have been offset to the mean sky coordinate for the sample, $(\alpha,\delta)=(184.7^\circ, 35.1^\circ)$  The spacing between grid lines is $10^\circ$ in latitude and $30^\circ$ in longitude.}
    \label{fig:sky-maps}
\end{center}
\end{figure*}

\begin{figure}
\begin{center}
\includegraphics[width=0.49\textwidth]{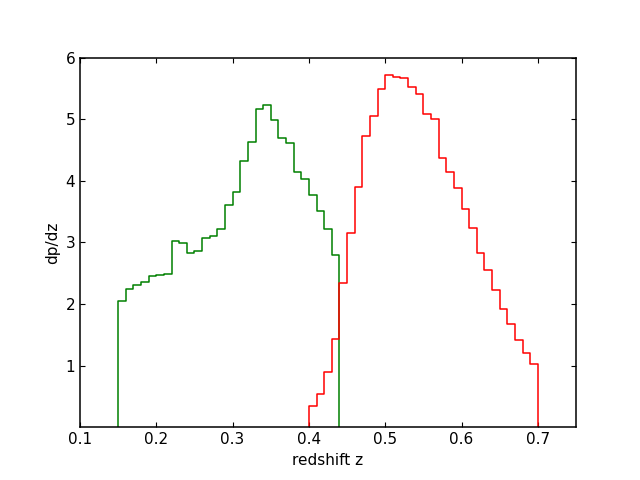}
    \caption{Redshift distributions for the two galaxy samples under study: LOWZ (green) and CMASS (red).
    }
    \label{fig:kernels}
\end{center}
\end{figure}

\begin{figure*}
\begin{center}
\includegraphics[width=0.49\textwidth]{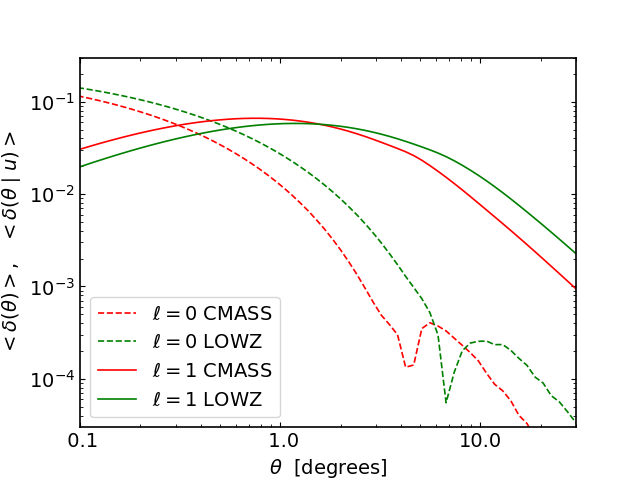}
\includegraphics[width=0.49\textwidth]{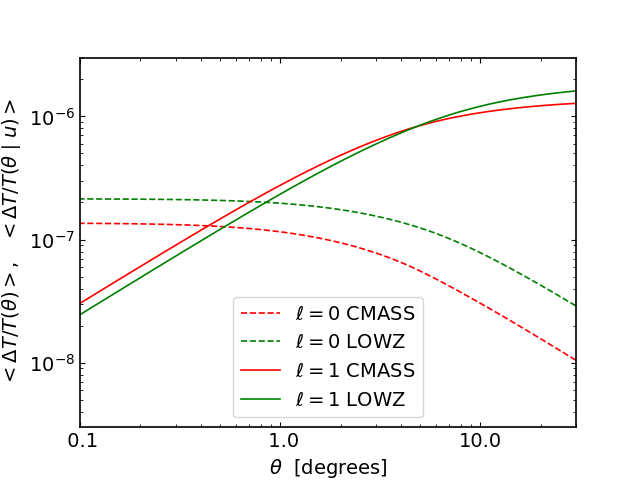}
\\
\includegraphics[width=0.49\textwidth]{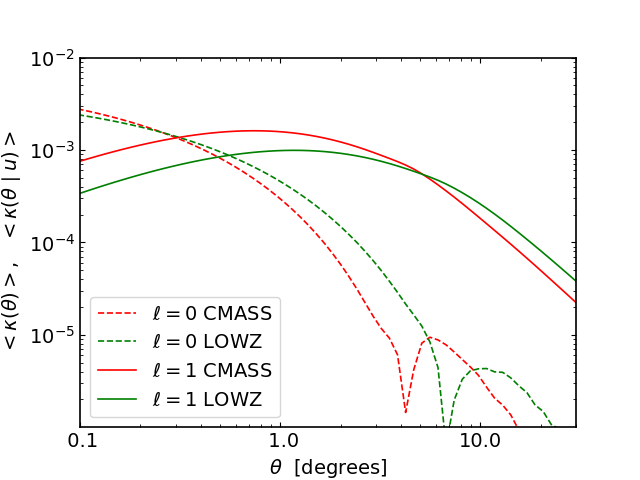}
\includegraphics[width=0.49\textwidth]{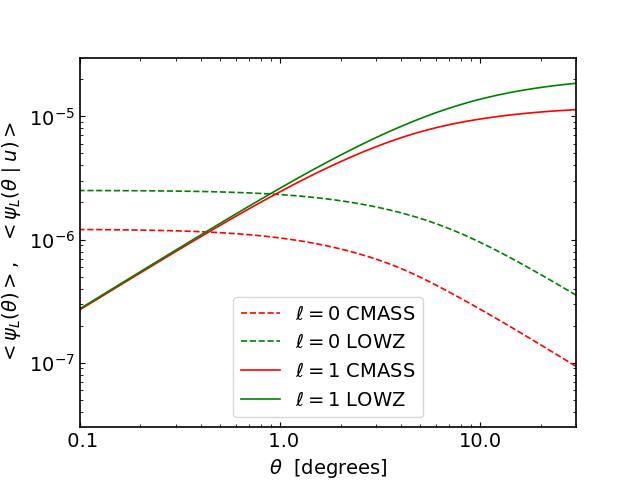}
    \caption{Predictions for the various angular multipoles of interest: projected overdensity; CMB temperature; CMB lensing convergence; CMB lensing potential. These are computed in the Limber approximation, using equations (\ref{eq:limber0}) and (\ref{eq:limber1}), together with (\ref{eq:dipole_amplitude}), as appropriate for velocity weighting of the dipole. Red lines denote CMASS, and green lines are for LOWZ. Solid lines are the dipole signal and dashed lines show the monopole. With the exception of overdensity, all dipoles are bias independent, while the monopoles gain a power of bias (there is an additional power of $b$ in the case of the density statistics). The appropriate values of $b$ have been applied to the predictions (2.01 for CMASS; 2.08 for LOWZ). For density and convergence, the predicted monopole is negative beyond roughly $5^\circ$; we plot the modulus.
    }
    \label{fig:limber}
\end{center}
\end{figure*}

\section{Observations of the dipoles}
\label{sec:observation}
To extract the dipole effects from observed data, we will use the transverse components of the velocity field reconstructed from galaxy redshift surveys. We then rotate and align the transverse velocity vectors of galaxies and stack using the CMB temperature map for the temperature dipole, and using the CMB lensing convergence map for the lensing dipole. Note that by choosing galaxies as the centres for stacking, we will also have non-zero monopoles, i.e. the conventional cross-correlation functions between the galaxy number density field with other fields. We emphasise again that the stacked dipole signal depends only the directions of the velocities, and not their amplitude. We use the public BOSS-DR12 LOWZ and CMASS galaxy samples, together with the PR3 release of the CMB temperature map and lensing convergence map from {\it Planck\/} for this analysis.

\subsection{BOSS galaxy data and velocity reconstruction}
We use data from the Sloan Digital Sky Survey \cite[SDSS;][]{York2000}, specifically data release 12 \cite[DR12;][]{Reid2016,Alam2014}. SDSS I, II \citep{Abazajian2009} and III \citep{Eisenstein2011} employed a drift-scanning mosaic CCD camera \citep{Gunn1998} on the 2.5m Sloan Telescope at the Apache Point Observatory in New Mexico \citep{Gunn2006} to image 14\,555 deg$^2$ of sky across five photometric bands \citep{Fukugita1996,Smith2002,Doi2010}, achieving a limiting magnitude of $r <22.5$.  The acquired imaging data underwent processing through various SDSS pipelines \citep{Lupton1999,Pier2003,Padmanabhan2008}. In Data Release 8 (DR8), \cite{Aihara2011} reprocessed all previous SDSS imaging data. 

The Baryon Oscillation Spectroscopic Survey \cite[BOSS;][]{Dawson2013} was conceived to obtain spectra and redshifts for 1.35 million galaxies across 10\,000 deg$^2$ of sky. These galaxies were selected from the SDSS DR8 imaging. For targeting in BOSS, an adaptive tiling algorithm developed by \citep{Blanton2003b} was employed, adjusting to the density of targets in the sky. BOSS utilised double-armed spectrographs \citep{Smee2013} for spectral acquisition; the redshift extraction algorithm used in BOSS is detailed in \citet{Bolton2012}. The survey achieved a high redshift completeness exceeding 97\% across the entire survey footprint, resulting in a homogeneous dataset.  An overview of the survey is provided by \citet{Eisenstein2011}, while \citet{Dawson2013} offers a comprehensive description of the survey design. 

From this observational material, we use the LOWZ and CMASS galaxy samples  \citep{Bolton2012} from DR12. Both of these samples consist of Luminous Red Galaxies (LRGs), which act as efficient LSS tracers over large volumes. We use CMASS galaxies between $0.4 < z < 0.7$ and LOWZ galaxies between $0.15 < z < 0.44$. The sky distributions of the galaxies in these samples are shown in Fig. \ref{fig:sky-maps}, and the corresponding redshift distributions in Fig. \ref{fig:kernels}.

For the present study, we need to estimate the peculiar velocities of individual galaxies; we achieve this by means of the following steps.

\begin{itemize}
    \item {\bf Coordinate Transformation:} We take the observed coordinates (i.e. RA, DEC, $z$) and convert them to comoving Cartesian coordinates assuming a flat $\Lambda$CDM fiducial cosmology.
    \item {\bf Density field estimation:} We use a 512$^3$ grid to estimate the density of galaxies using the cloud in cell (CIC) scheme \citep{Hockney1981}. We use $50 \times$ randoms to calculate the selection function on the grid. The overdensity is calculated by taking the ratio of galaxy and random counts and subtracting unity. Whenever the randoms have zero density the overdensity is also set to zero.
    \item {\bf Smoothing the density field:} We first divide the overdensity by the linear bias for both samples \citep{2016MNRAS.457.1770C} to convert this to the overdensity field of the underlying mass distribution. This is then smoothed with a Gaussian kernel of scale $\sigma_{\rm smooth}=10\mpcoh$. 
    \item {\bf Displacement field:} We solve for the \cite{Zeldovich} displacement field using a standard multigrid technique\footnote{\url{https://github.com/martinjameswhite/recon_code/blob/master/notes.pdf}}. We used the following estimates of the growth rates at the mean redshift: $\fg=0.68$ for LOWZ and $\fg=0.77$ for CMASS for the purpose of estimating the displacement field. This gives the displaced position of each of the galaxies in our sample. 
    \item {\bf Velocity Estimates:} The displacement vector gives the velocity when multiplied by $H(z) \Omega_m(z)^{0.55}/(1+z)$. This is then projected into components along the line-of-sight of sight and tangential to  the line-of-sight. We emphasise again that the absolute amplitudes of the velocities do not affect our analyses -- only the direction matters.
\end{itemize}

The reconstruction code used in this process is publicly available\footnote{\url{https://github.com/martinjameswhite/recon_code}} and uses the standard lowest order algorithm described in \cite{Eisenstein07}.

The BOSS catalogues are known to contain some systematics, resulting from e.g. contamination of the LRG candidates by stars. To account for those, we use the column called {\rm WEIGHT$_{\rm SYSTOT}$} provided in the DR12 LSS catalogue. This includes the total systematic weight for each galaxy to account for imaging systematics, fibre collisions, and redshift failures \citep{2017MNRAS.464.1168R}. Each galaxy (or redshift) is weighted by the supplied weight when generating the map of galaxy number counts and when carrying out stacking.

For the masks of the LOWZ and CMASS samples, we project the randoms into {\sc Healpix} sky pixels \citep{Gorski2005}, recording unity if there is at least one random per pixel, and zero otherwise. Since the randoms have high density, 50 times higher than the actual galaxy sample, and the highest resolution we have used for sky maps is with {\sc Nside\,=\,512}, this method is sufficient. 

Being Luminous Red Galaxies, both the LOWZ and CMASS samples are strongly biased. The linear bias parameters for these samples were determined in a detailed 3D clustering analysis by \cite{2017MNRAS.465.1757G}: $b=2.01$ (CMASS) and $b=2.08$ (LOWZ). These values will be assumed in interpreting the clustering multipole measurements that we present below. Knowing the bias and having the $N(z)$ distributions, we can then predict the ISW and lensing multipoles for these samples. We use the Limber formalism set out in Section \ref{sec:limber_prediction}. The result is shown in Fig. \ref{fig:limber}.

\begin{figure*}
\begin{center}
 \scalebox{1.0}{
 \includegraphics[width=1.0\textwidth]{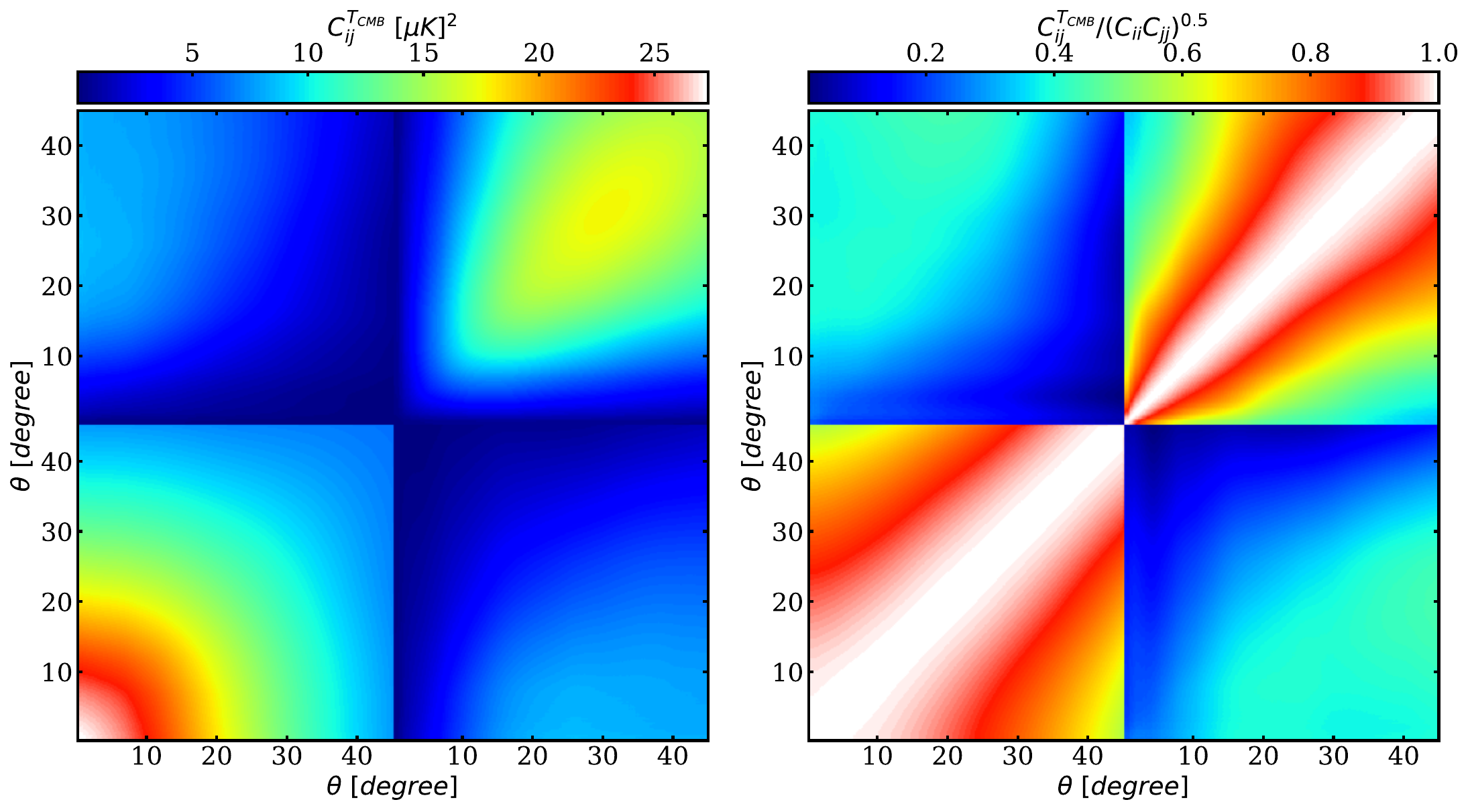}}
    \caption{Left: the covariance matrix for the monopole (bottom left) and dipole (top right) of the stacked CMB temperature constructed from repeating the stacking with 500 random-phase CMB realisations. Right: The correlation coefficients of the matrix on the left. The correlations between the monopole and dipole are weak as expected, and so the dipole should provide extra cosmological information in addition to the monopole.}
    \label{fig:stacked_Tcmb_Cov}
\end{center}
\end{figure*}

\begin{figure*}
\begin{center}
 \scalebox{1.0}{
 \includegraphics[width=1.0\textwidth]{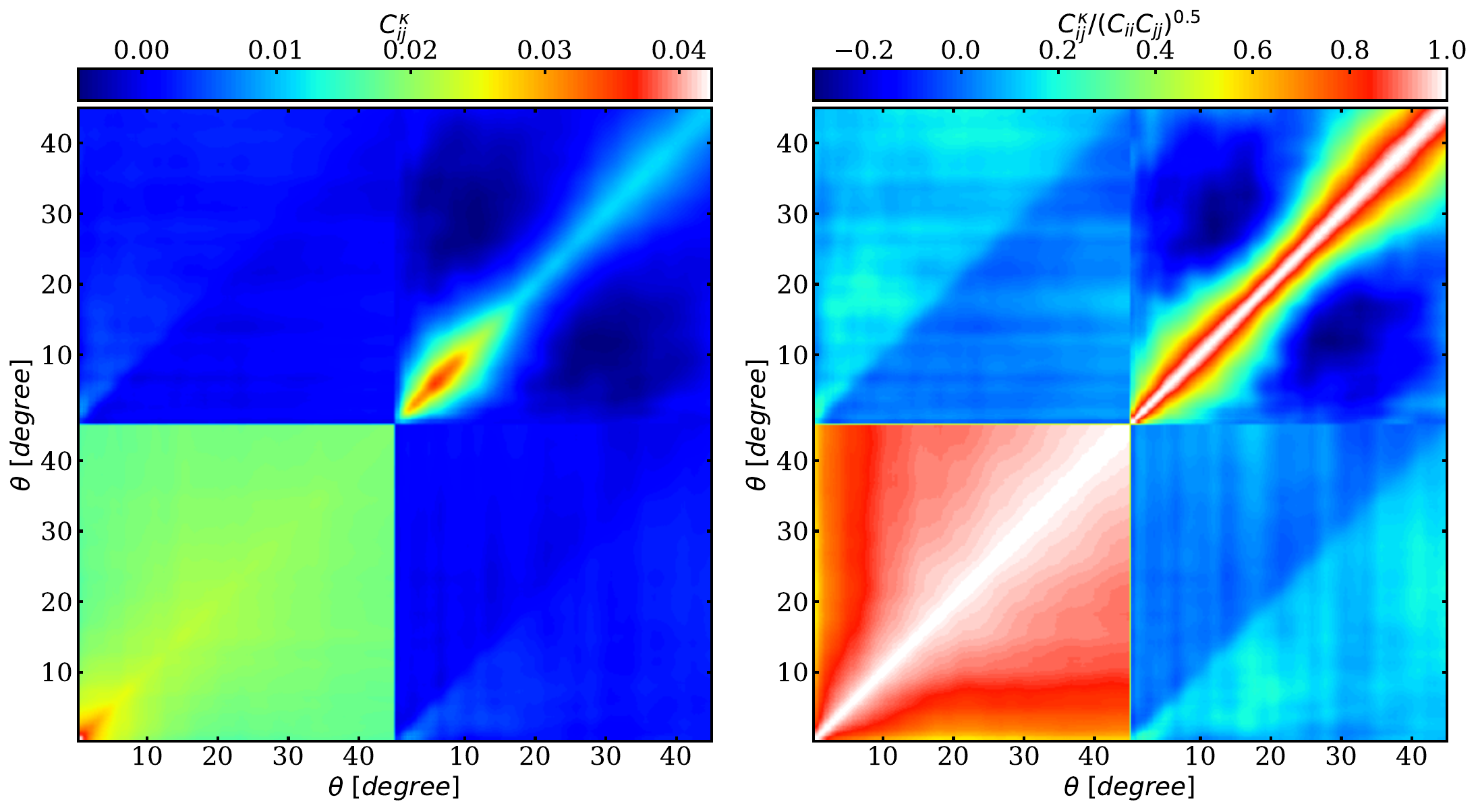}}
    \caption{Left: the covariance matrix for the monopole (bottom left) and dipole (top right) of the stacked CMB lensing convergence $\kappa$, constructed from repeating the stacking with 500 random-phase CMB realisations. Right: The correlation coefficients of the matrix on the left. The correlations between the monopole and dipole are weak as expected, and so the dipole should provide extra cosmological information in addition to the monopole. The dipole is also more diagonal than the monopole.}
    \label{fig:stacked_kappa_Cov}
\end{center}
\end{figure*}

\subsection{CMB temperature and lensing data}

We use CMB temperature and lensing convergence maps from the 2018 PR3 release of the {\it Planck\/} satellite \citep{Planck2018_overview, Planck2018_lensing}\footnote{\url{http://pla.esac.esa.int/pla}}. The PR3 lensing dataset supplies harmonic information up to $\ell=4096$, although in practice we imposed a maximum wavenumber of $\ell=2048$ in creating the convergence map.
We also use the {\it Planck\/} PR2 estimated map of the ISW effect, which was derived using information from the distribution of galaxies inferred from several galaxy surveys of large-scale structure \citep{PlanckISW2016}. The exact datasets used are listed in Table \ref{tab:datasets}.

\begin{table}
\caption{Specific {\it Planck\/} CMB datasets used in this analysis.}
\begin{center}
\begin{tabular}{c} 
\hline
Filename \\
\hline
{\tt COM\_CMB\_IQU-commander\_2048\_R3.00\_full.fits} \\
{\tt COM\_Lensing\_4096\_R3.00.tgz}\\
{\tt COM\_CompMap\_ISW\_0064\_R2.00.fits}\\
\hline
\end{tabular}
\end{center}
\label{tab:datasets}
\end{table}

\subsection{Dipole stacking methodology}

For each galaxy with its sky coordinate $\vec r$ and reconstructed velocity vector $\vec v$, we can extract its velocity component on the sky by evaluating
\begin{equation}
\vec v_{\perp}  = \hat{\vec r} \times (\vec v \times \hat {\vec  r}),
\end{equation}
where $\hat {\vec  r}$ is the unit vector for the position of the galaxy. The stacking is performed with two steps of rotation:

(1) We have maps of $\delta_g({\vec \theta})$, $T_{\japsc CMB}(\vec \theta)$ and $\kappa_{\japsc CMB}(\vec \theta)$, representing the projected galaxy number density contrast, CMB temperature, and CMB lensing convergence, respectively. Each of these is given a shifted centre using the sky coordinates of each target galaxy, so that the target point becomes the north pole in the new map.  

(2) We then rotate each of the background maps in azimuth around the new centre so that the transverse velocity for each galaxy lies along the horizontal direction, pointing from left to right. The shifted and rotated maps can now be stacked.

To take care of the masks, we repeat the above stacking with the masks for the CMB temperature map, lensing convergence map, and the survey footprints of the BOSS galaxy sample respectively, $M(\vec \theta)$. The end product of the stacking is the mask-weighted average of all maps:
\begin{equation}
S^J_{\rm stack} (\vec \theta)  = \frac{ \sum_{i=1}^{N} {w_iM_i^J(\vec \theta) S_i^J(\vec \theta) }}{\sum_{i=1}^{N} {w_iM_i^J(\vec \theta)}},
\end{equation}
where $J=1, 2, 3$ represent quantities of the galaxy number density contrast, CMB temperature map and lensing convergence map, respectively; $w_i$ is the systematic weight for the $i^{th}$ galaxy.

Finally, we decompose the stacked 2D map into a monopole and a dipole using Legendre polynomials:
\begin{equation}\label{eq:decomposition}
S^J_{\ell}(\theta) = \frac{(1+2\ell)}{2}\int_{-1}^{1} S^J(\theta, \mu)P_{\ell}(\mu)  \,d\mu, 
\end{equation}
where $P_{\ell}(\mu)$ are the Legendre polynomials of with $\ell$ being an integer; $\mu = \cos\vartheta $ and $\vartheta$ is the subtended angle from the velocity vector on our 2D map. $S^J_{\ell}(\theta)$ with $\ell=0, 1$ are the monopoles and dipoles.

\begin{figure*}
\begin{center}
\scalebox{1.0}{
	\includegraphics[width=0.49\textwidth]{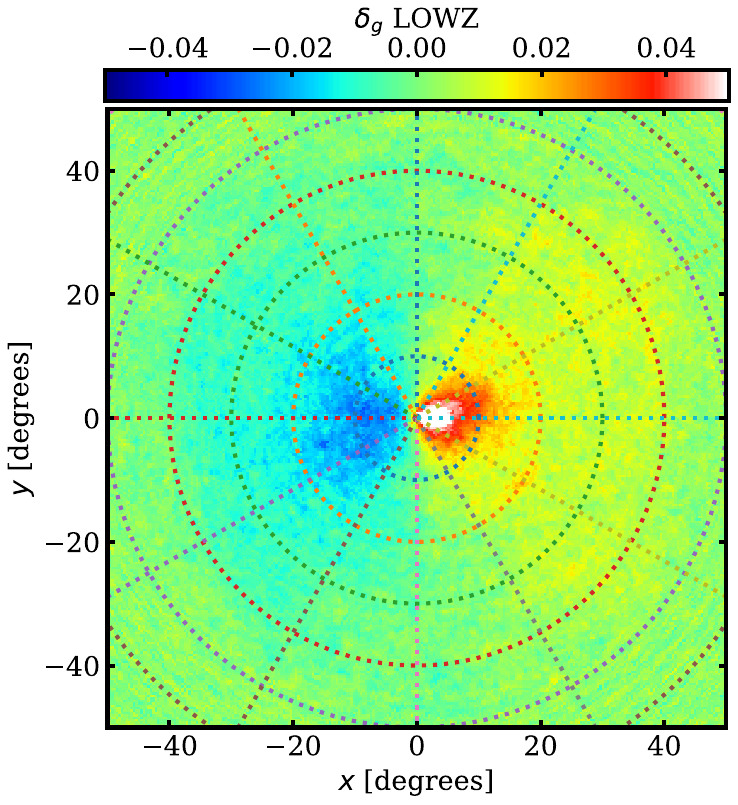}}
 \scalebox{1.0}{
 \includegraphics[width=0.49\textwidth]{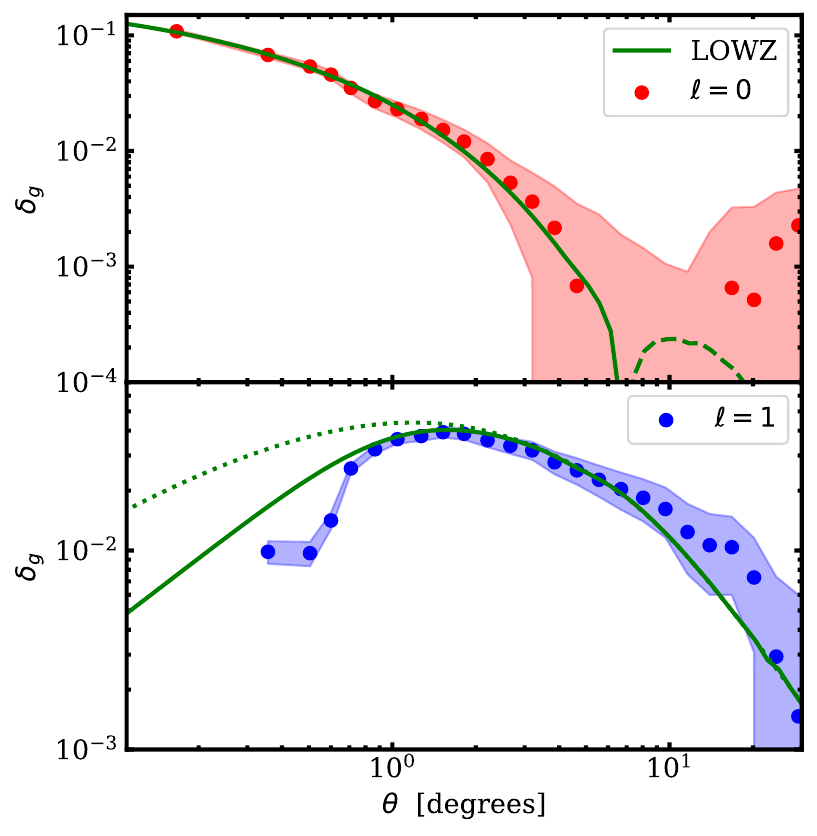}}
\scalebox{1.0}{
	\includegraphics[width=0.49\textwidth]{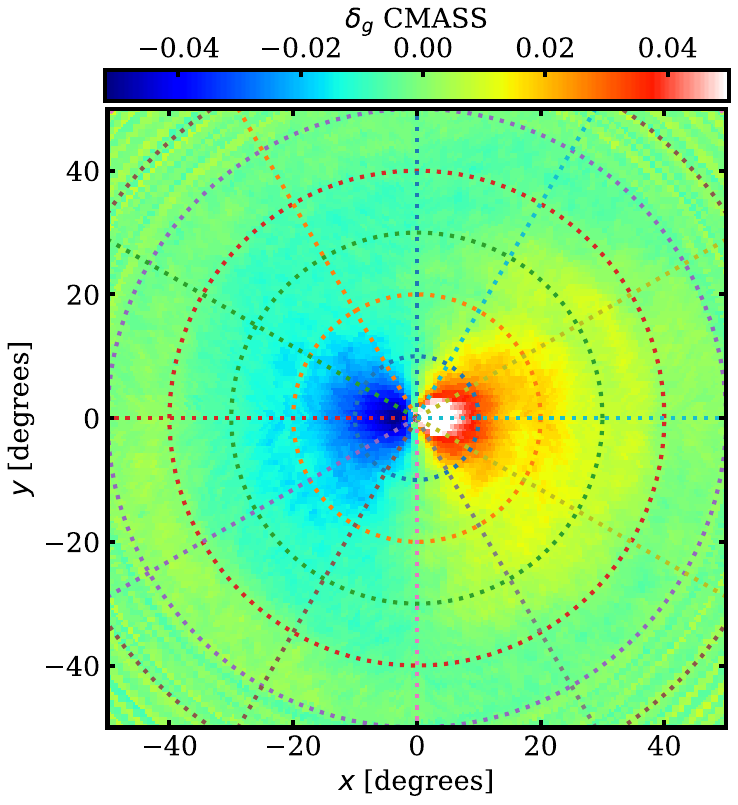}}
 \scalebox{1.0}{
 \includegraphics[width=0.49\textwidth]{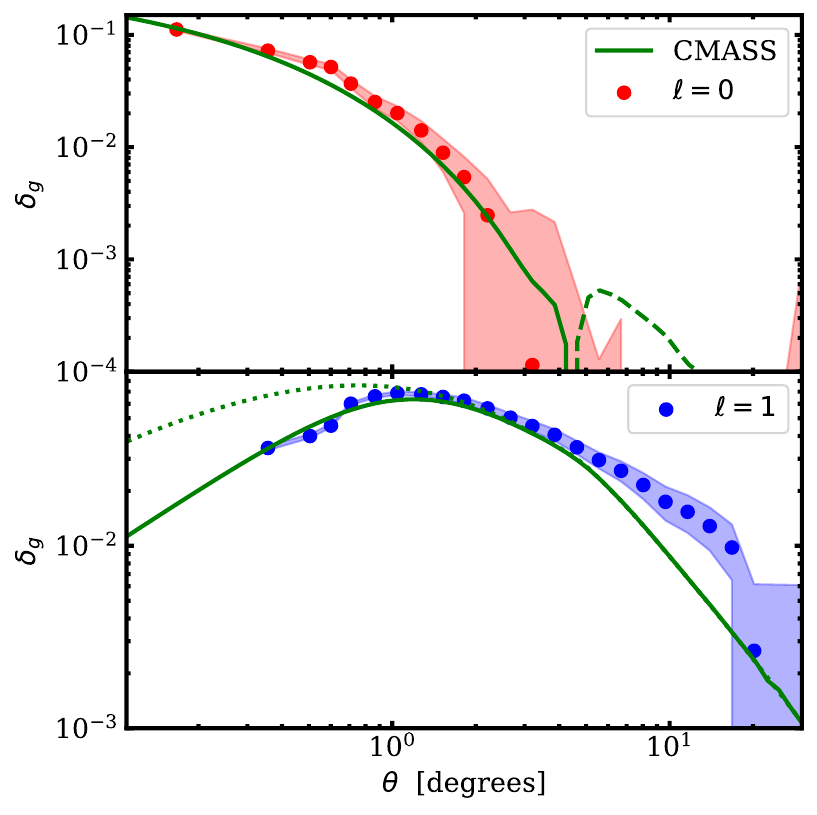}} 
    \caption{Azimuthal equidistant projection of the results from stacking maps of galaxy number density contrast $\delta_g$ at the location of galaxies in the BOSS LOWZ and CMASS samples. Maps are shifted and rotated such that galaxies are in the centre of the stack. The maps are aligned according to the reconstructed transverse velocities, with the velocity vector pointing from left to right, and averaged using the velocity amplitude as a weight.
    Left: Azimuthal equidistant projection of the stacked $\delta_g$ map. Right: monopole and dipole of the stacked $\delta_g$ map; shaded regions are the 1-$\sigma$ errors computed from 500 QPM mocks \citep{White2014}. The green curves are monopoles and dipoles predicted from the fiducial cosmology, with $b=2.08$ and $b=2.01$ for the LOWZ and CMASS samples, respectively.     
    The monopole is essentially the conventional projected galaxy correlation function; the dipole profile quantifies the amplitude of the dipole as a function of scale. Dashed lines in the monopole plots represent negative values. For the dipoles, solid and dotted lines represent predictions with and without allowance for the window function used in the velocity reconstruction (see the main text for details). }
    \label{fig:density_multipoles}
\end{center}
\end{figure*}

\begin{figure*}
\begin{center}
\scalebox{1.0}{
	\includegraphics[width=0.49\textwidth]{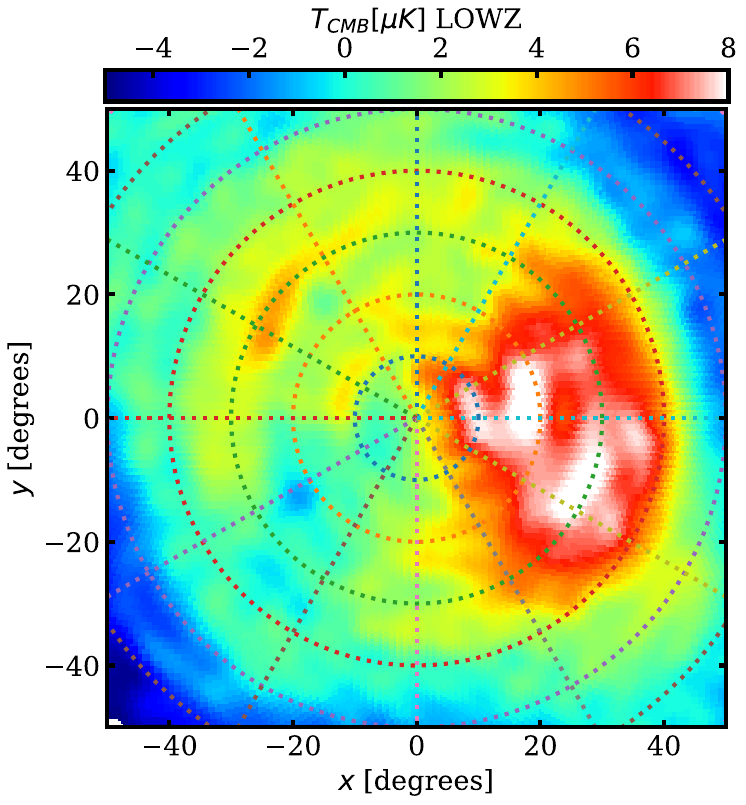}}  
 \scalebox{1.0}{
 \includegraphics[width=0.49\textwidth]{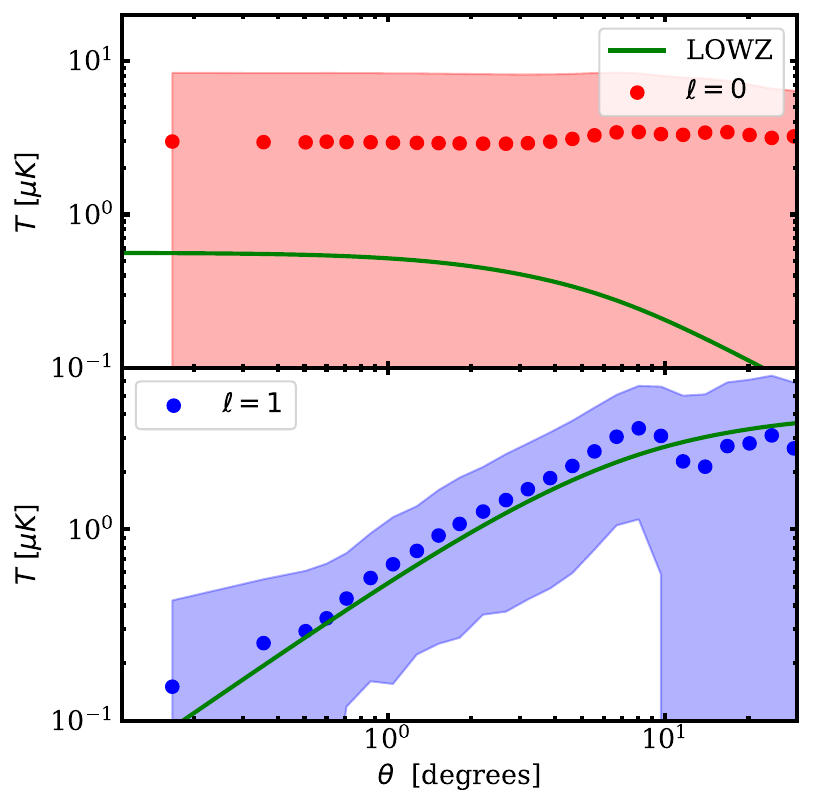}
 }
\scalebox{1.0}{
	\includegraphics[width=0.49\textwidth]{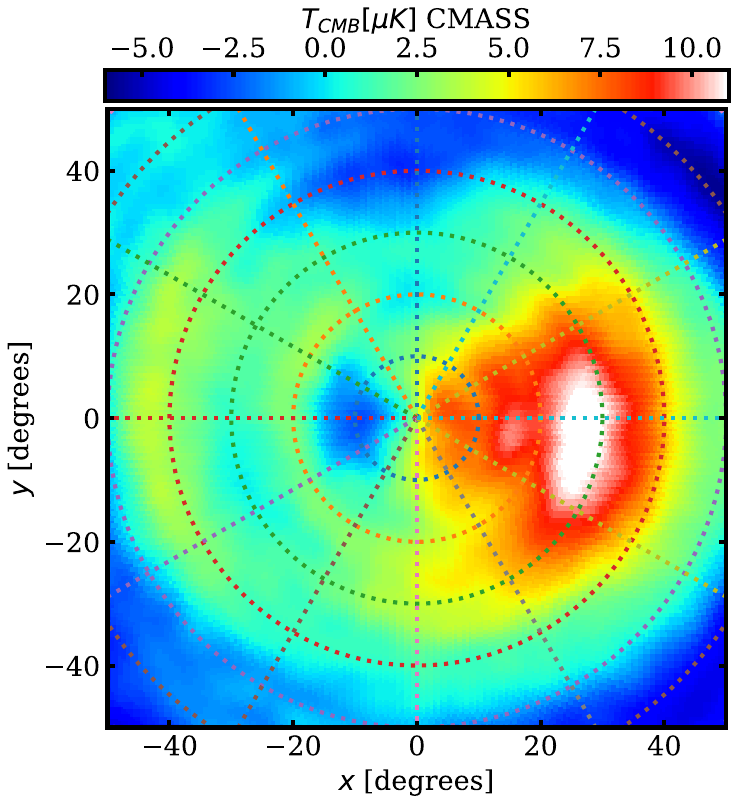}}
 \scalebox{1.0}{
 \includegraphics[width=0.49\textwidth]{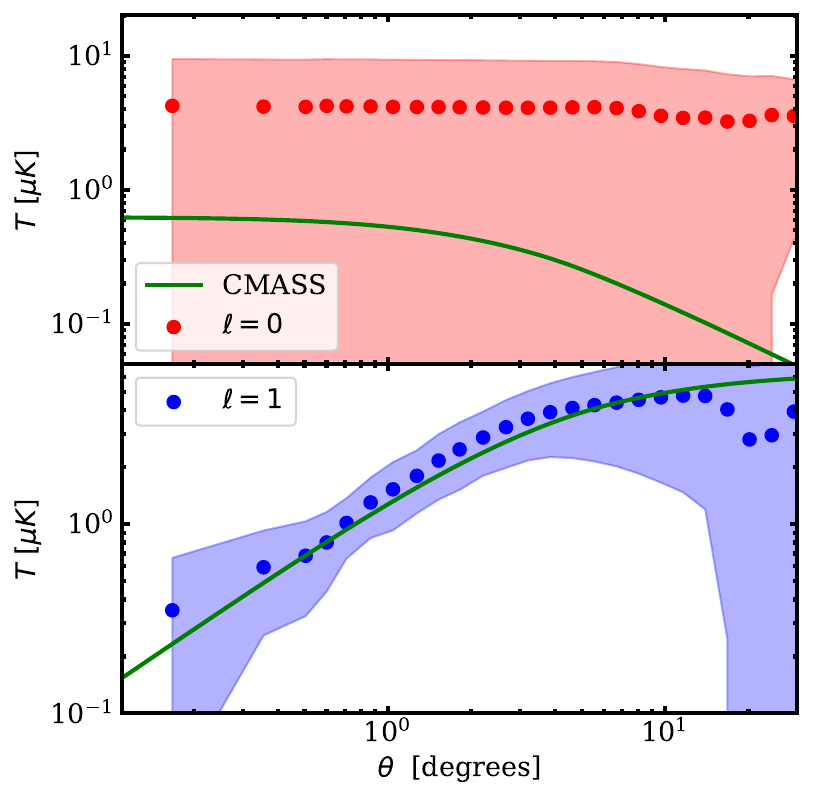}
 }
    \caption{Left: Azimuthal equidistant projection of the results from stacking the {\it Planck\/} CMB temperature map at the location of galaxies in the LOWZ and CMASS samples. Maps are shifted and rotated such that galaxies are in the centre of the stack. The maps are aligned according to the reconstructed transverse velocities, with the velocity vector pointing from left to right, and averaged using the velocity amplitude as a weight.
    Left: Azimuthal equidistant projection of the stacked {\it Planck\/} CMB temperature map. Right: monopole and dipole of the stacked temperature map; shaded regions are the 1-$\sigma$ errors from the diagonal component of the covariance matrix shown in Fig.~\ref{fig:stacked_Tcmb_Cov}. The green curves are the best-fit models for the LOWZ and CMASS samples, respectively. These fits were derived using the dipole signal only, and they incorporate the best-fit dipole amplitude parameters given in Table~\ref{table:amplitudes}.  The monopole is essentially the averaged CMB temperature profile around CMASS galaxies, i.e. the conventional $\delta_g$-$T_{\rm CMB}$ cross-correlation function; the dipole profile quantifies the amplitude of the dipole as a function of scale.}
    \label{fig:stacked_Tcmb_CMASS}
\end{center}
\end{figure*}

\begin{figure*}
\begin{center}
\includegraphics[width=0.49\textwidth]{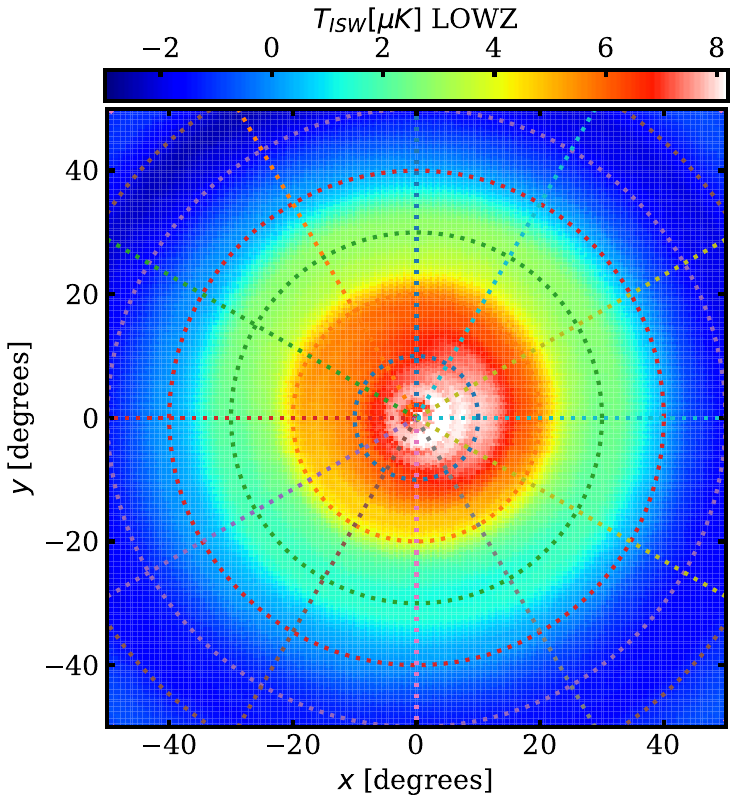}
\includegraphics[width=0.49\textwidth]{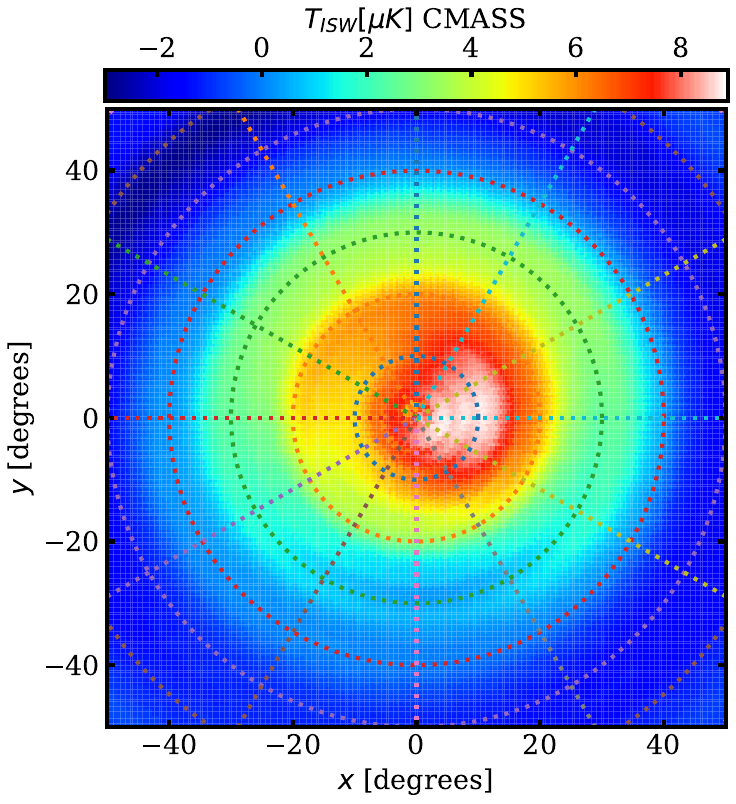}
\caption{Azimuthal equidistant projection of the stacked reconstructed ISW temperature map \citep{PlanckISW2016} with galaxies from the LOWZ and CMASS samples. The maps are shifted and rotated such that the position of galaxies are in the centre. The maps are aligned according to the reconstructed transverse velocities, with the velocity vector pointing from left to right.}
    \label{fig:stacked_ISW}
\end{center}
\end{figure*}

\subsection{Error estimation}

In judging the significance of any mismatch between the
measured and predicted multipole signals, we need to understand the
precision of the data; this is normally quantified by a covariance
matrix, so that we can use a Gaussian likelihood. The covariance
can be estimated in two standard ways: either by means of random
data realisations, or via an internal sampling strategy such as the
jackknife. We have applied both of these approaches. Generation of
random CMB skies is straightforward, and the variance from the
fluctuations in the primary temperature fluctuations is heavily
dominant over the small ISW signal. Similarly, the lensing map is
accurately Gaussian, and it is easy to create realisations
using the total noise+signal power spectrum of the empirical lensing
map. We experimented with also making lognormal realisations
of the galaxy data, but the uncertainties from the CMB dominate.
The only issue is that these realisations have no true multipole
signal built into them -- but given that the S/N of the multipole
detections are modest, we felt that it was unnecessary to create
realisations that included foreground effects at the predicted level.
We compared these simple direct covariance estimates with a
jackknife, based on dividing the survey region into 160 subregions.

The covariance estimates agree reasonably well, and examples are presented in Figs \ref{fig:stacked_Tcmb_Cov} and \ref{fig:stacked_kappa_Cov}, which show respectively the realisation-based covariance of the temperature multipoles and the CMB lensing convergence multipoles. These plots demonstrate that there is relatively little statistical coupling between the monopole and dipole, as expected -- so they do indeed supply nearly independent cosmological information. It is also noticeable that the off-diagonal covariance elements of the dipoles are significantly smaller than those of the monopoles, especially for the lensing convergence. Thus the data points in the dipoles are relatively more independent from each other, which gives an advantage when we seek to extract information from the dipole signal.
However, the covariance plots also display the common issue that the data values for each multipole are very highly
correlated amongst themselves: this means that the formal Gaussian likelihood approach
is problematic, with the result tending to be dominated by the small
differences between adjacent data values. This is in addition to
the concern that even inevitable small imperfections in the covariance
could lead to important errors in its inverse. We therefore settled on
a simpler procedure. We identified a core angular range where our
results should be affected neither by map resolution nor by large-angle
effects in the theory, and simply fitted the theory to the data over this
range by least squares, deducing a multipole amplitude scaling factor, $A$. This same process was applied to the data
realisations, yielding a direct measure of the uncertainty in $A$.
We used an angular range of 1 -- 10 degrees, fitting to the
data with a uniform sampling in $\log(\theta)$. The details of
these choices had no important impact on the deduced amplitude
scalings and their precision.

\begin{figure*}
\begin{center}
\scalebox{1.0}{
	\includegraphics[width=0.49\textwidth]{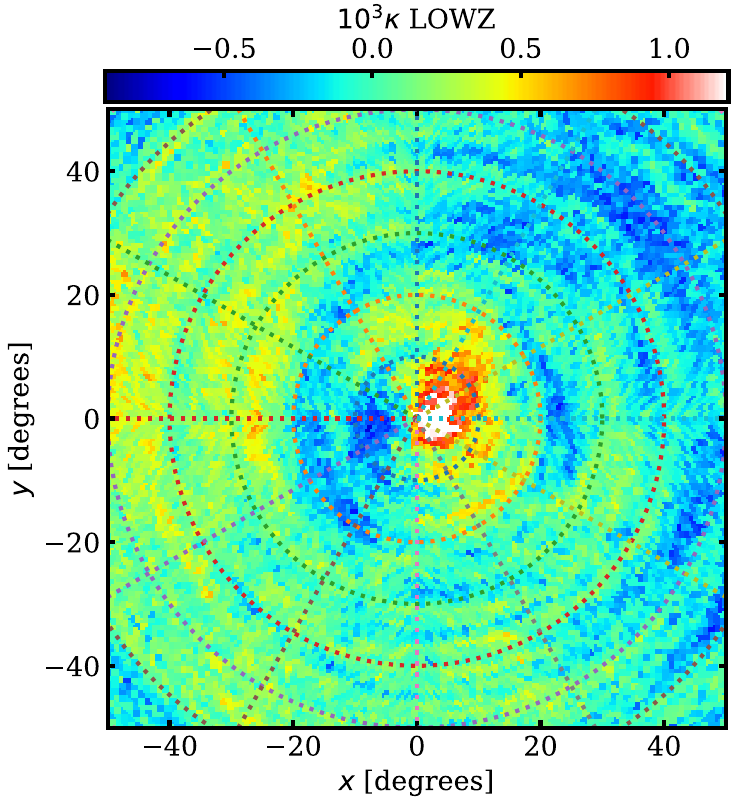}}
 \scalebox{1.0}{
 \includegraphics[width=0.49\textwidth]{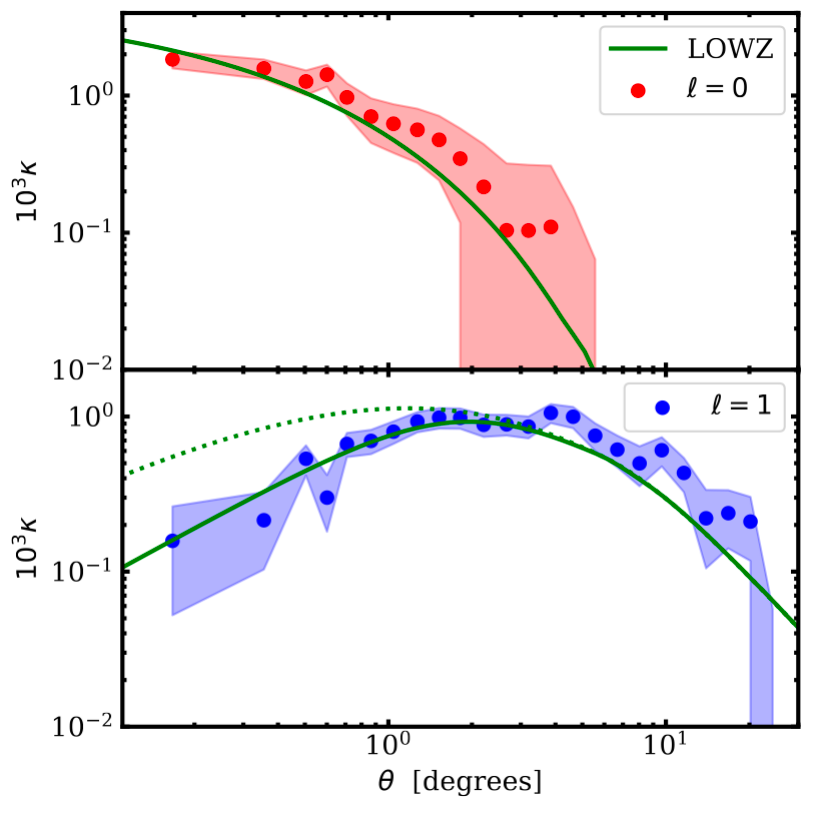}}
\scalebox{1.0}{
	\includegraphics[width=0.49\textwidth]{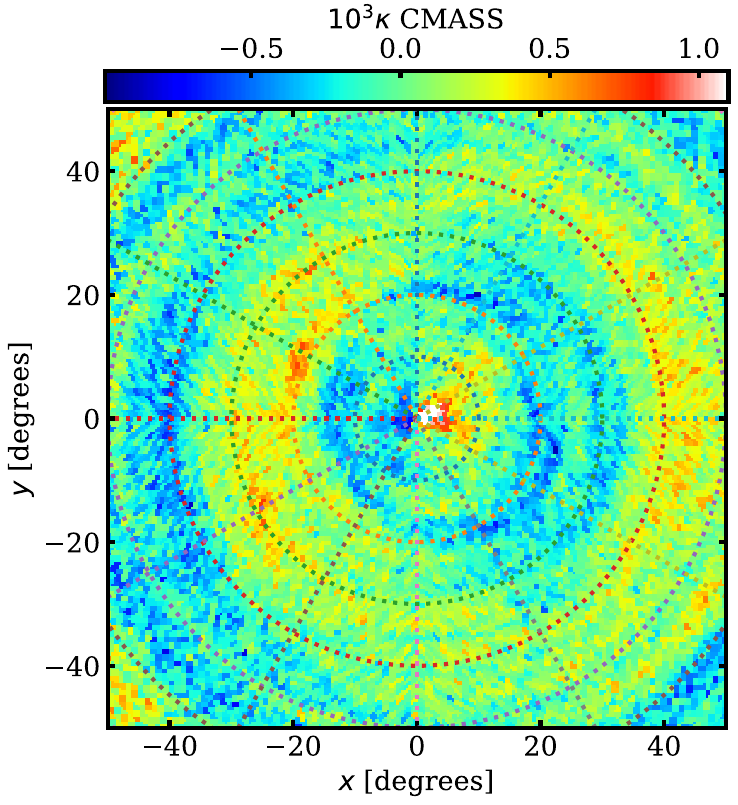}}
 \scalebox{1.0}{
 \includegraphics[width=0.49\textwidth]{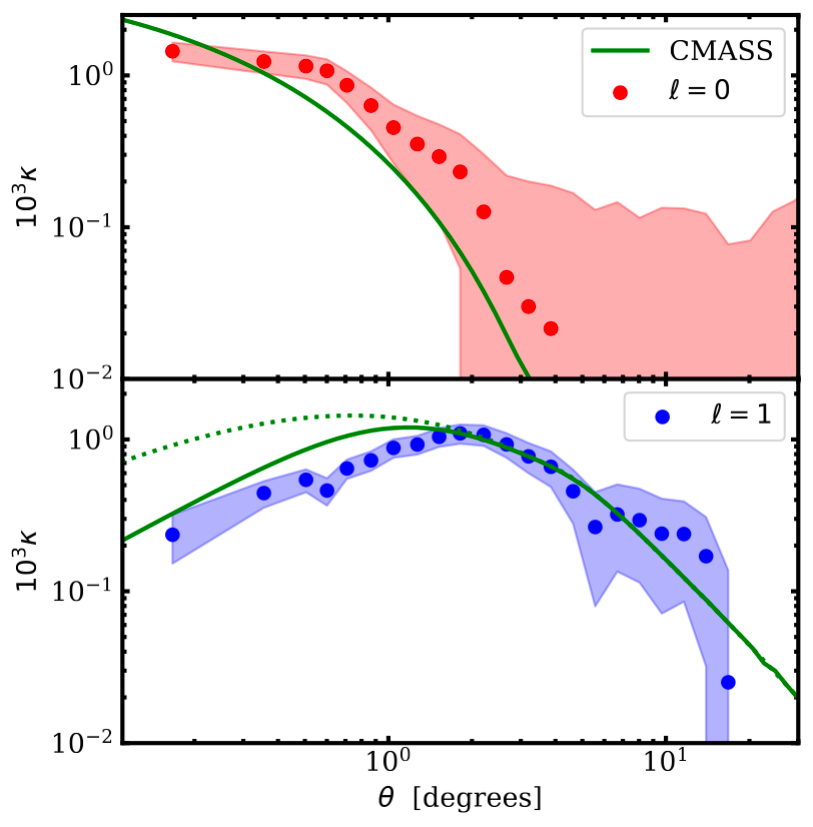}}
    \caption{Similar to Fig.~\ref{fig:stacked_Tcmb_CMASS} but showing the results from stacking the CMB lensing convergence map from {\it Planck\/}. For the dipoles, solid and dotted lines represent predictions with and without allowance for the window function used in the velocity reconstruction (see the main text for details). }
    \label{fig:stacked_kappa}
\end{center}
\end{figure*}

\begin{figure*}
\begin{center}
\scalebox{1.0}{	\includegraphics[width=0.487\textwidth]{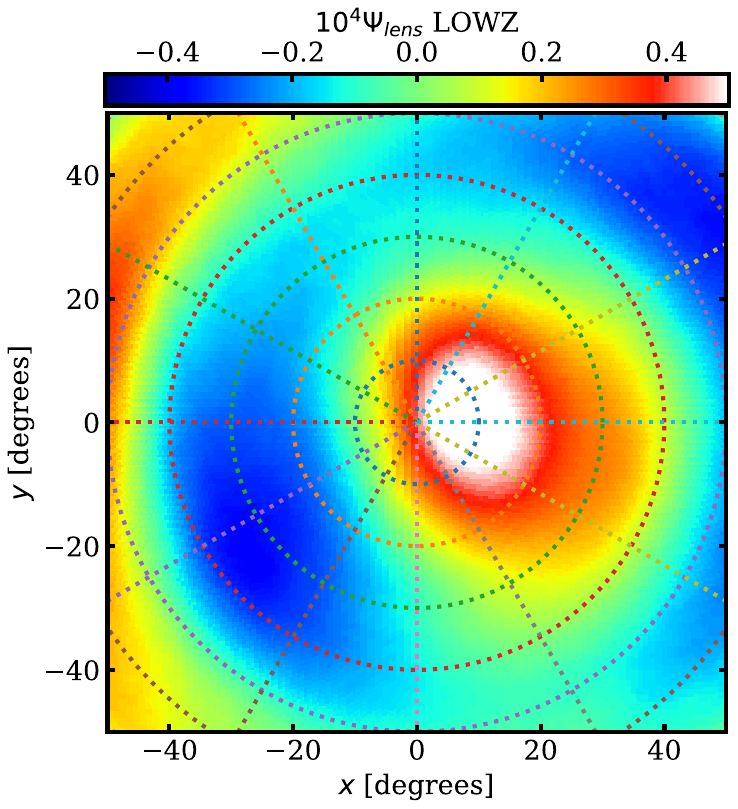}}
\scalebox{1.0}{	
\includegraphics[width=0.498\textwidth]{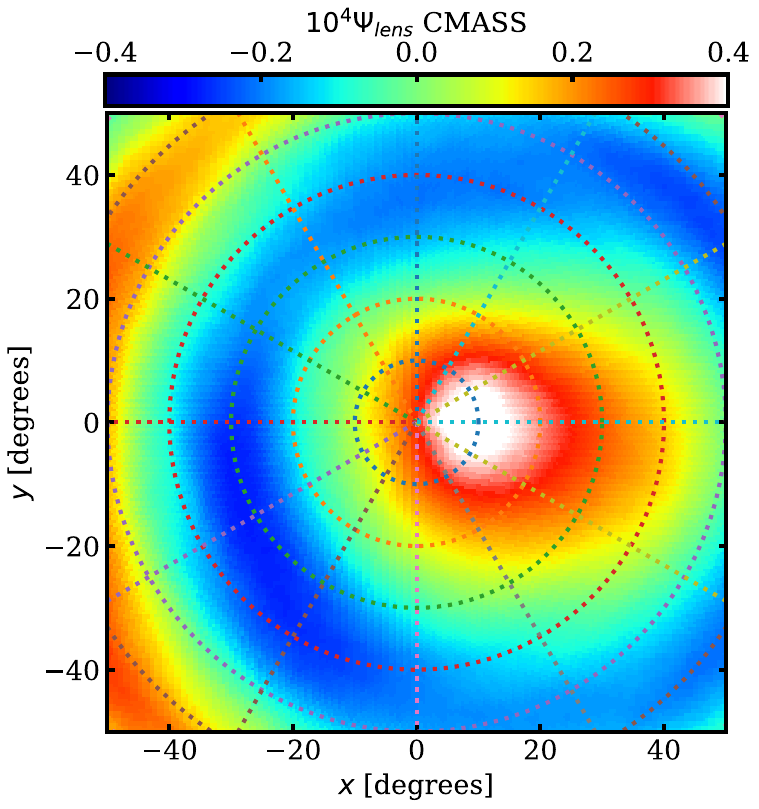}}
    \caption{Azimuthal equidistant projection of the stacked {\it Planck\/} CMB lensing potential map $\psi_{\rm Lens}$ with galaxies from the LOWZ and CMASS samples. The maps are shifted and rotated such that the position of galaxies are in the centre. The maps are aligned according to the reconstructed transverse velocities, with the velocity vector pointing from left to right. Modes with $\ell<5$ are set to zero to suppress spurious large-scale fluctuations due to the mask.}
    \label{fig:stacked_psi}
\end{center}
\end{figure*}

\section{Results}
\label{sec:results}

\subsection{The density dipole}

We first perform stacking of the galaxy overdensity field. The monopole signal
here will be simply the angular 2-point correlation function, $w(\theta)$, but
the dipole in the stacked density will be a novel signature. This is an important consistency test, since the peculiar velocities that we use for the stacking are inferred from the galaxy density field. It would
therefore be most surprising if the galaxy density field failed to show a
dipole that aligns with the velocity. However, this exercise does serve as
a test both of the apparatus for predicting the multipole signals, and of the
quality of the data. Regarding the dipole in particular, one might imagine
that in principle the velocity reconstruction could have noise in direction
that would lead to the amplitude of the dipole being systematically low; it
is therefore a useful test to see if the density dipole is recovered as
predicted. We note that these predictions require a knowledge of the
bias for the galaxy sample under study, with the monopole scaling as $b^2$
and the dipole as $b$. We use standard figures for these, as discussed earlier.
To estimate the errors on these measurements, we repeat our measurement with 500 mock galaxy catalogues generated using the `quick particle mesh' (QPM) method described in \cite{White2014}. Their 1-$\sigma$ errors are shown as the shaded regions, but we do not use these for our cosmological constraints.

Figure \ref{fig:density_multipoles} shows the density multipoles, which generally display satisfactory
agreement with prediction at small scales. But on scales of order 10 degrees
the agreement is less good. In particular, the monopole is expected to fall
to zero at about $5^\circ$ (CMASS) or $7^\circ$ (LOWZ), whereas the data
show a continued positive signal at these separations -- especially for
CMASS. The galaxy density maps incorporate the official systematic weights
that attempt to allow for known sources of non-uniformity in the galaxy sample,
but it seems that this process is not completely successful. This is not
entirely surprising, since it is well-known that the large-scale correlations
measured in the CMASS spectroscopic sample are not in accord with linear
theory. BAO analyses of the CMASS sample therefore incorporate additional
empirical `broad-band' terms in the modelling. For the present purpose, we
see that the surface density map must contain spurious large-scale
variations. These can be eliminated by subtracting a filtered version
of the data from the map, with the filtering scale chosen so that the
large-scale monopole falls to zero as expected. In practice, a Gaussian
filter of FWHM $30^\circ$ achieves the desired result, and we apply this
process to both LOWZ and CMASS maps. This in effect supplies an additional
position-dependent weight to be applied to the galaxy samples, with an
rms of a few percent. This weighting makes little change to the density dipole,
but we apply it in all our analyses.

On scales below about $1^\circ$, the density dipoles appear to be lower than theoretical expectation. This is mainly due to the smoothing we have applied in reconstructing the velocity vectors, and the possible non-linearity of the velocity on small scales. As shown below, we see similar behaviour for the dipole of the lensing convergence, for the same reasons. However, the generally good agreement at larger separations between linear theory and observation for these multipoles of galaxy number density gives us confidence in performing the same analyses with the CMB temperature and lensing dipoles. We will define multipole amplitudes using scales $1^\circ<\theta<10^\circ$, in order to avoid effects of small-scale velocity smoothing and large-angle effects in our theoretical models.

\subsection{The temperature dipole}
The left-hand panel of Fig.~\ref{fig:stacked_Tcmb_CMASS} presents the results from the cross-correlations between the BOSS transverse velocity samples and the {\it Planck\/} CMB temperature map. To suppress potential systematic effects on large scales, we use modes in the range $2<\ell<60$. The upper bound in $\ell$ has little impact on the scales of interest, but it helps to reduce the small-scale noise in the 2D map. A clear signal of the CMB temperature dipole is detected. The dipole extends beyond the scale of a few tens of degrees, and its direction aligns well with the transverse velocity vector (which  points from the left to the right). 

The dipole signal departs from zero more clearly than the monopole. This is consistent with the results of 
\cite{2014MNRAS.438.1724H}, who attempted to detect the ISW signal in cross-correlation with BOSS data and achieved only a marginal ($1.6\sigma$) detection of the ISW monopole.
We note that the monopole is subject to an additive correction based on the mean value of the map being stacked. For the CMB, the average temperature over the BOSS area departs from the global 2.725\,K as a result of the large-scale fluctuations in the temperature power spectrum. The question is then whether any offset in this empirical mean should be subtracted, so that the monopole is forced to zero at large angles. This is what was done e.g. by \cite{2014MNRAS.438.1724H}, and we take the same approach here. This step is somewhat ad hoc, but it reminds us that the observed monopole is heavily affected by the large-scale temperature modes, whereas the dipole is not.

As discussed further below, we believe that this dipole signal constitutes a clear detection of the ISW signal that arises within the BOSS survey region. As an immediate direct illustration that this is likely to be the correct interpretation, we can also stack the reconstructed ISW map that was created by the {\it Planck\/} team by combining temperature data with information on the galaxy distribution from several galaxy surveys of large-scale structure \citep{PlanckISW2016}.
The result is shown in Fig.~\ref{fig:stacked_ISW}. Because of the somewhat complicated data combination used in this case, we will not attempt to quantify the significance of the detection here, and will focus on the results from the CMB temperature stack, which should contain all the ISW signal contributed by large-scale structure. The agreement here between the modelled dipole and observational data does not show the small-scale deviation that we saw in the dipoles of galaxy number density and lensing convergence. This is mainly due to the large-scale coherence of the gravitational potential, which makes it less affected by the imperfection of the reconstructed peculiar velocities.

\subsection{The lensing dipole}
Figure \ref{fig:stacked_kappa} presents the results for the stacking with the CMB lensing convergence map. We use modes in the range $6<\ell<256$, in order to suppress noise from large scales. A dipole signal with $\kappa$ pointing along the horizontal direction is clearly detected. The effect appears to be concentrated towards smaller scales in comparison with the temperature dipole -- as expected because the $\kappa$ dipole reflects the projected mass fluctuation around the velocity vector, whereas an ISW temperature dipole relates to the gravitational potential.
There is also a clear monopole signal down to degree scales and below. 

If we were to translate this dipole into its corresponding lensing potential $\Psi_{\rm L}$, its coherence scale would be much larger. It should be similar to that of the temperature dipole, as they both arise from the projection of the gravitational potential along the line of sight. Converting $\kappa$ to the lensing potential $\Psi_{\rm L}$ can be done in harmonic space using $\Psi_{\rm L}(\ell)=2\kappa(\ell)/[\ell(\ell+1)]$. However, due to the incompleteness of the sky map, the mask of the $\kappa$ map will interfere with the large-scale modes, causing spurious large-scale fluctuations in the resultant map of $\Psi_{\rm L}$. To mitigate this effect, we cut off modes with $\ell<5$: the resulting lensing potential stack is presented in Fig.~\ref{fig:stacked_psi}. We can see that the dipole is indeed smoother and has a larger coherence scale than the convergence map, despite the removal of the low-$\ell$ modes. We caution that the pattern of the map changes with different cuts of low-$\ell$ modes.

\begin{figure*}
\begin{center}
\includegraphics[width=0.49\textwidth]{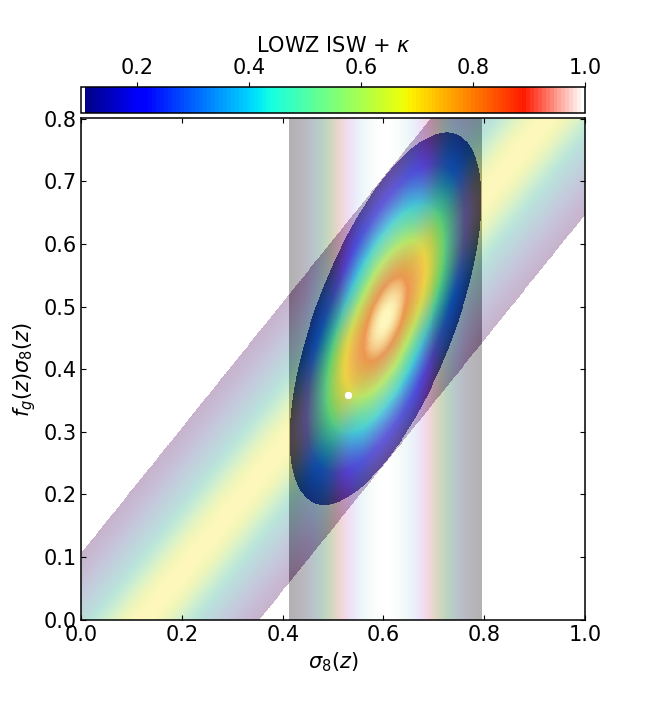}
\includegraphics[width=0.49\textwidth]{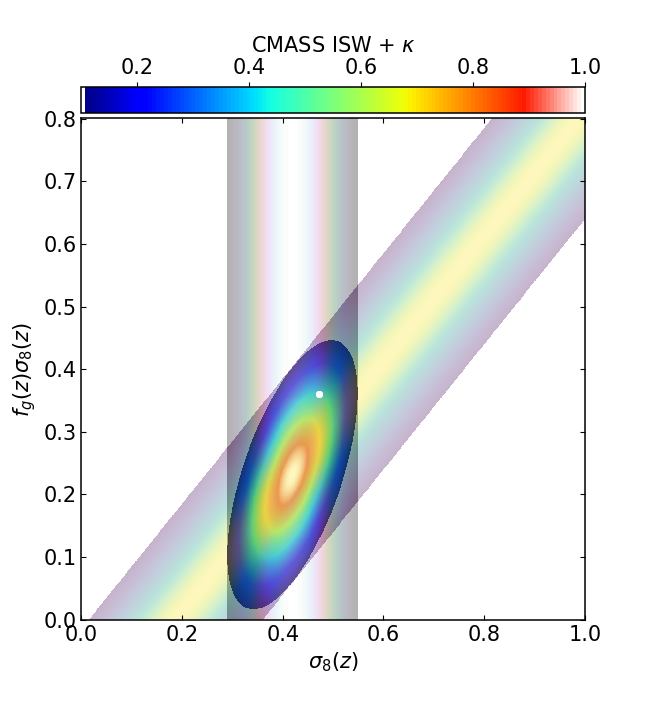}
    \caption{Constraints on the plane of $\fg(z)\sigma_8(z)$ from the combination of the dipoles in CMB lensing $\kappa$ and CMB temperature, interpreted as the ISW effect. The main colour coding shows the combined likelihood relative to the maximum, with separate $\kappa$ and ISW results illustrated semi-transparently. The $\kappa$ measurement constrains only $\sigma_8(z)$, while ISW constrains a diagonal band. Results are shown separately for LOWZ and CMASS. The white point marks the prediction of the fiducial model. 
    }
    \label{fig:fsig}
\end{center}
\end{figure*}

\begin{figure*}
\begin{center}
\includegraphics[width=0.49\textwidth]{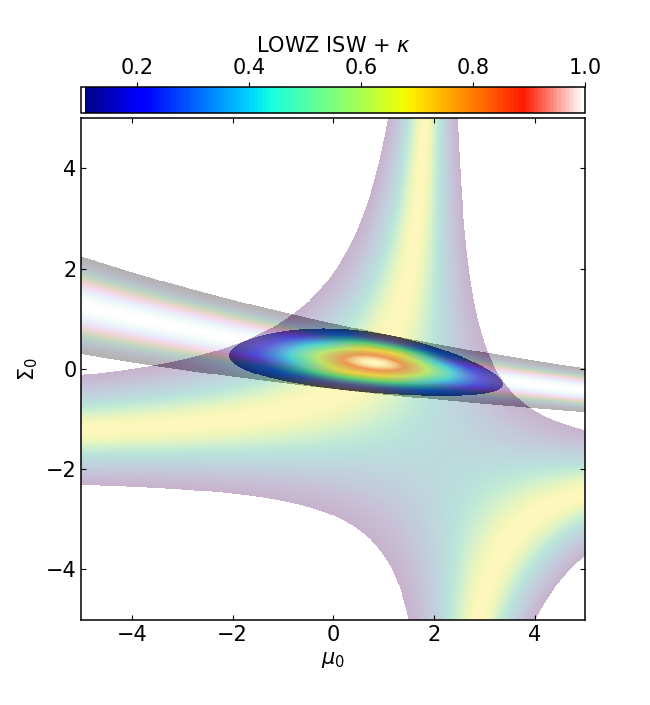}
\includegraphics[width=0.49\textwidth]{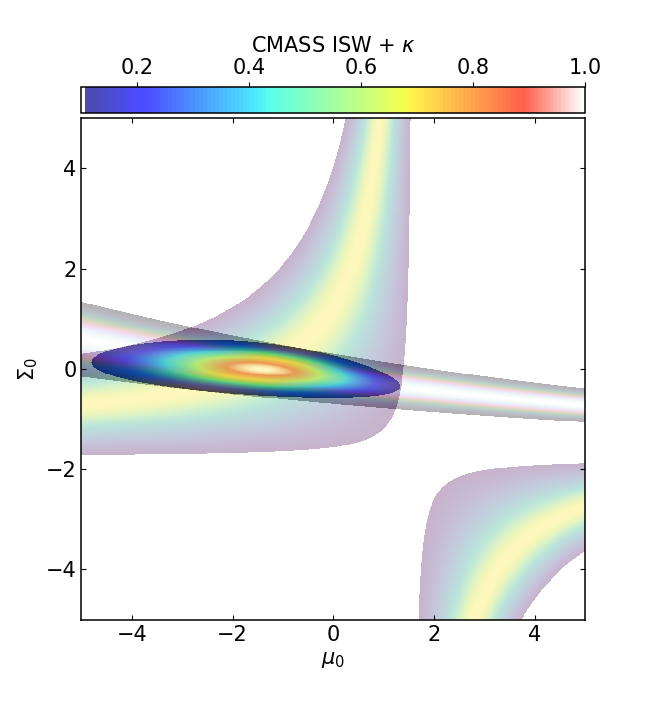}
    \caption{Constraints on the modified gravity parameters $\mu_0$ and $\Sigma_0$ from the combination of the dipoles in CMB lensing $\kappa$ and CMB temperature, interpreted as the ISW effect. he colour coding is the same as in Fig. \ref{fig:fsig}. Results are shown separately for LOWZ and CMASS.
    }
    \label{fig:mg}
\end{center}
\end{figure*}

\section{Dipole Modelling: growth and implications for gravity}
\label{sec:interpretation}

The interpretation of the multipole amplitudes depends in principle on all the parameters of the cosmological model. But there would be little point in performing a full exploration of that parameter space, as many parameters would be unconstrained or constrained only in degenerate combinations. If we bring in external data from CMB, BAO, and SNe, this will have the effect of fixing very accurately the parameters that constrain the expansion history, cosmic geometry, and amplitude of density fluctuations, and so we should already have accurate predictions for the ISW and CMB lensing signals within $\Lambda$CDM. But these predictions need not be correct in non-standard theories of gravity, and this is the possibility we explore here. External data will fix the empirical expansion history, which we know is close to $\Lambda$CDM with $\Omega_m=0.3$, but the amplitudes of the ISW and lensing signals will depend in addition on the growth of fluctuations. Thus we can ask whether the values of $\fg$ and $\sigma_8$ are consistent with the fiducial values. We can do this simply by introducing amplitude parameters in which the ISW and lensing signals are scaled by factors $A_T$ and $A_\kappa$ with respect to the fiducial predictions, as discussed above.
For lensing, a dipole amplitude $A\neq 1$ tells us that $\sigma_8(z)$ differs from the fiducial value, so that we have an estimator $\hat\sigma_8(z) = A\,\sigma_{8,\,\rm fid}(z)$. Similarly, the ISW dipole depends on $\sigma_8(z)[1-\fg(z)]$. The monopoles depend on one extra power of $b(z)\sigma_8(z)$; this factor is constrained empirically by the observed clustering of the target galaxies, so the interpretation of the monopole scaling would be the same as for the dipole. However, since the dipoles are the novel element of this paper, we now focus purely on their interpretation, i.e. we exclude the monopole in our model fitting. We do achieve agreement between our best-fit model and monopole data, for both CMB temperature (Fig.~\ref{fig:stacked_Tcmb_CMASS}) and CMB lensing (Fig.~\ref{fig:stacked_kappa}), even though the model is determined solely by constraints from the dipoles. This agreement strongly indicates the consistency of our modelling, as well as of our observational measurements, for both the monopoles and the dipoles.

\begin{table}
\caption{Amplitude scaling factors for the lensing and temperature dipoles, relative to the fiducial cosmology prediction.}
\begin{center}
\begin{tabular}{ccc} 
\hline
Sample & $A_\kappa$ & $A_T$ \\
\hline
LOWZ & $1.14\pm0.17$ & $0.72\pm0.63$ \\
CMASS & $0.89\pm0.13$ & $1.69\pm0.73$ \\
\hline
\label{table:amplitudes}
\end{tabular}
\end{center}
\end{table}

We present in Table~\ref{table:amplitudes} the derived lensing and temperature dipole scalings, with their errors, for the LOWZ and CMASS samples. We now comment on the extent to which the results are consistent with $A=1$.
Constraints on the $\fg-\sigma_8$ plane from this approach are shown in Fig. \ref{fig:fsig}. We see that the large observed temperature dipole amplitude for CMASS drives us towards either a high value of $\sigma_8$ or a low value of $\fg$. In contrast, the LOWZ results are more closely consistent with the fiducial prediction. 

But this empirical approach of allowing the amplitude and growth rate of structure to vary freely is undesirable in two ways. First of all, the amplitude is an integral over the growth rate. But more importantly, in modified gravity models it is possible or even likely that the two metric potentials $\Psi$ and $\Phi$ will differ from each other, and both lensing and ISW signals are sensitive to this possibility, being proportional to $\Psi+\Phi$. We therefore approach the interpretation of our dipole data through an explicit modification of gravity, following standard approaches in which this is characterised by two parameters -- which in effect alter the gravitational constant, $G$, and the `slip' relation between the potentials. There are a variety of ways in which this approach can be notated, but we follow \cite{Simpson2013} in defining parameters that are most closely related to observed signals:
\[
\eqalign{
\Psi &= (1+\mu)\,\Psi_{\rm GR} \cr
\Psi + \Phi &= (1+\Sigma)\, [\Psi + \Phi]_{\rm GR}.
}
\]
This says that, for a given density fluctuation, $\delta$, the forces on non-relativistic particles change by a factor $1+\mu$ compared to standard expectations based on $\delta$, while the forces on relativistic particles alter by a factor $1+\Sigma$. The offsets $\mu$ and $\Sigma$ will in principle change with redshift, and it is normal to assume that they become negligible at high $z$: this preserves the standard interpretation of the CMB if the expansion history is fixed, but it is also justified by the idea that the late-time accelerated cosmic expansion is related in some way to the modification of gravity. Again we follow Simpson et al.:
\[
(\mu,\Sigma)=(\mu_0,\Sigma_0)\, {\Omega_\Lambda(z)\over \Omega_\Lambda},
\]
which assumes that the evolution of $(\mu,\Sigma)$ follows that of the dark energy density parameter in the $\Lambda$CDM model.
An issue with this parameterisation is that it does not tell us directly about modifications of $\sigma_8$ or $\fg$. For this, we have to solve numerically the linear growth equation for $\delta$, given that the effective $G$ is boosted by a factor $1+\mu$:
\[
\ddot \delta + 2{\dot a\over a}\, \dot\delta = (1+\mu)4\pi G \bar\rho(a)\, \delta
=(1+\mu)\, {3\over 2} H_0^2\Omega_m a^{-3} \delta.
\]
A practical form of this is to define $D\equiv \delta/a$ and to use dashes to denote $d/da$:
\[
\eqalign{
D'' + &\left({5\over a}+{H'\over H}\right)\, D' +
\left({3\over a^2}+{H'\over aH}\right)\, D
= \cr &(1+\mu)\, {3\over 2} \left({H_0\over H}\right)^2 \Omega_m a^{-5} D,
}
\]
where we have $H^2(a)=H_0^2(1-\Omega_m + \Omega_m a^{-3})$. We set $D=1$ and $D'=0$ at some small initial value of $a$ and integrate to the era of interest. This yields $\delta$ and $\fg=d\ln\delta/d\ln a$ as a function of $\mu_0$. Finally, we insert these modified values in the expression for the lensing and ISW effects and multiply by
$1+\Sigma$.

The dipole measurements then give constraints on the $\mu_0-\Sigma_0$ plane, as shown in Fig. \ref{fig:mg}. The ISW measurement shows an interesting pattern, in which $\Sigma_0$ can be positive or strongly negative, so that the sign of $\Psi+\Phi$ flips; but this can be compensated by a large value of $\mu_0$, which yields a growth rate $\fg>1$. In contrast, the lensing dipole largely yields a constraint on $\Sigma_0$. The combination of these two constraints shows a weak tension with $(\mu_0,\Sigma_0)=(0,0)$ for each of LOWZ and CMASS, but in rather different directions, so that overall the datasets appear to be in good agreement with standard gravity.

\section{Discussion and conclusions}
\label{sec:summary}

In this paper we have made a systematic exploration of the ways in which the transverse peculiar velocity field can be used as a signpost for foreground effects in cosmology, arising from large-scale structure along the line of sight. The key is to exploit the coupling between velocities and their underlying gravitational potentials in the linear regime. We thus predict the presence of large-scale dipoles in galaxy number density, gravitational lensing and the CMB ISW effect, all aligned with transverse velocities. 

The transverse velocities are not directly observable, but can be inferred from the galaxy distribution -- robustly in terms of direction, and to within an amplitude scaling set by the galaxy linear bias. This inference depends on the assumption that density fluctuations are purely in an irrotational growing mode at early times. In this case, and in the linear regime, the velocity is deduced from the dipole moment of the density field, with an appropriate radial weighting.

We have shown that this means of inferring peculiar velocities can be tested and exploited, because a variety of gravitational effects can be expected to have a dependence on direction that leads them to align with the peculiar velocity (i.e. with the density dipole). In the earlier sections of the paper we assembled the necessary theory for these effects, and then set out to see if they could be detected. The tool for achieving this is rotational stacking. We take astronomical maps that should contain a dipole signal that aligns with the local transverse velocity, and extract postage stamps around a number of stacking centres -- which we then rotate individually in order to align the velocities prior to stacking. In this way, we have explored a number of distinct effects, providing a set of novel probes of the cosmological model:

\renewcommand{\theenumi}{(\arabic{enumi})}
\begin{enumerate}
\smallskip
\item We use the LOWZ and CMASS datasets from the SDSS-III BOSS survey as our source of the velocity field, making use of the fact that this is routinely estimated via the displacement field in order to carry out BAO reconstruction. This provides thick slices of the galaxy density and velocity fields, centred at redshifts of $\langle z \rangle\simeq 0.3$ and $\langle z \rangle\simeq 0.5$, respectively. This galaxy surface density field contains a dipole of the expected amplitude, demonstrating the consistency of the velocity reconstruction, and the effectiveness of the stacking strategy.

\smallskip
\item We then consider the two gravitational effects on photons coming from the CMB, the ISW effect and CMB lensing, using the temperature maps and CMB lensing maps from {\it Planck\/}. In the linear regime, both effects are proportional to line integrals of the gravitational potential, and both are expected to display a dipole that aligns with the transverse velocity. Our rotational stacking allows both these effects to be detected at a high significance. Importantly, these dipoles are an independent signal from the standard lensing and ISW monopoles, and so offer extra cosmological information. The amplitudes of the signals are close to the prediction of the fiducial $\Lambda$CDM model. We have used these measurements to set limits on models of modified gravity.  

\smallskip
\item We also discuss the nonlinear effect of the `moving gravitational lens'. This was first analysed by \cite{Birkinshaw1983}, who pointed out that there should be an induced dipole in CMB radiation that passes the lens. This dipole was originally claimed to be in the sense that the peculiar velocity would point to a temperature enhancement. But we confirm previous suggestions that this result was in error, and that the correct effect is that the velocity points to a temperature decrement. 

\smallskip
\item 
We show that the ISW dipole is distinct from the well-studied dipole induced by the moving-lens effect \citep{Rubino-Martin2004, Cai2010, Yasini2019, Hotinli2021, Beheshti2024}. They are both aligned with the transverse velocity, but have opposite signs for the temperature, with the larger-scale ISW dipole having the velocity point towards a temperature enhancement. So the full picture should be that the small-scale ($\sim$\,$10\mpcoh$) moving-lens dipole is embedded in a large-scale ($\sim$\,$100\mpcoh$) ISW dipole that has the opposite sign. One way of seeing the sign of the moving-lens effect is to present it as an aspect of the Integrated Sachs-Wolfe effect. Ahead of the moving lens, the potential well is clearly deepening with time, so that $\dot\Phi<0$. But this is the opposite sign from the linear ISW effect, which is dominated by the fact that the potential wells causing the peculiar velocities are decaying as $\Lambda$ becomes dominant. In the linear regime, $\dot\Phi(\vec{x}) \propto \Phi(\vec{x})$, and so there should be a dipole in the linear ISW effect that aligns with the peculiar velocity.

\smallskip
\item We have demonstrated that the dipole signal has a range of statistical advantages: (i) the dipole is independent from the monopole, so this signal can therefore provide additional cosmological information in addition to the monopole. The density-velocity coupling relationship is set only when a model of gravity is assumed, and so the velocity field provides extra information in constraining gravitational degrees of freedom; (ii) the data vector of the dipole shows weaker correlations than the monopole, boosting the overall S/N of the measurement; (iii) The dipoles of gravitational potential, and related quantities, are independent of galaxy bias -- an attractive feature for a probe of the large-scale structure, which is also found in gravitational lensing. 

\smallskip
The dipole signal thus lies
beyond what was conventionally considered as two-point statistics, i.e. the two-point correlation function or the power spectrum. In our terminology, these represent only the monopole signal.  We therefore see this study of velocity-related dipole effects as a first exploration of a powerful new cosmological signature. This has yielded encouraging results, in particular a substantial improvement in the significance of detection of the ISW effect. 
Improvements in the analysis are certainly possible, as we have taken the simplest approach of using dipoles alone in the model fitting: a full exploitation of cosmological information using both the monopole and the dipole should see improved constraints on cosmological parameters, especially when applied to new galaxy datasets such as DESI. The large-scale nature of the gravitational dipoles we have detected suggest that they could be promising probes of primordial non-Gaussianity and effects of general relativity that are typically prominent on very large scales \citep[e.g.][]{Yoo2009, Bonvin2011, Challinor2011}. Finally, there are additional foreground effects that could be studied, and extensions to the theory to give a full treatment of small-scale non-linearities will be of interest. The initial results presented here give strong reason to believe that further efforts in this direction should be fruitful.

\end{enumerate}

\section*{Data availability}
The BOSS LOWZ and CMASS galaxy samples and maps of CMB temperature and CMB lensing from {\it Planck} are public. The simulations used in this paper, plus our codes and results produced, are available upon reasonable request to the authors.

\section*{Acknowledgments}
We thank Alex Hall and Andy Taylor for useful discussions, and Carlos Hern\'andez-Monteagudo for a very helpful referee's report. YC acknowledges the support of the UK Royal Society through a University Research Fellowship. YC is grateful for the hospitality of the Astrophysics and Theoretical Physics groups of the Department of Physics at the Norwegian University of Science and Technology during his visit, when part of this work was conducted. For the purpose of open access, the author has applied a Creative Commons Attribution
(CC BY) license to any Author Accepted Manuscript version arising from this submission.
\appendix

\section{measurements from simulations}
\label{appA}
Given the theoretical setup, we can compare the predictions with measurements from N-body simulations. Since the observables probe the velocity field, they are sensitive to the largest perturbations modes due to the $1/k$ factor in Fourier $k$-space for the velocity. We need to use large volumes of simulations to beat down cosmic variance at the very large scales. To do this, we employ the simulations created by \cite{Alam2017}. This suite of simulations was run assuming a flat $\Lambda$CDM universe with $\Omega_m=0.292$ in 10 independent cubic boxes, each of side $1380\mpcoh$. This paper presents results from simulations using the average of the 10 boxes.

Figure~\ref{v-correlation-1D} compares the velocity correlations between linear theory predictions with simulations. We can see that they agree very well and well within the errors for all scales. 

Figure~\ref{v-correlation-2D} is the 2D version of the stacked velocity field. To measure this from simulations, we assign particles to cubic grids of $200^3$ using the CIC scheme \citep{Hockney1981}, and calculate the averaged density and velocity on the grids. We then treat the $z$-axis as the LOS direction. We then average the velocity field $\vec v(x,y)$ along the $z$ direction over the whole length of the box, i.e. projecting out the $z$ component. On the $x$-$y$ plane, we have the averaged velocity fields $\vec v(x,y)$. For stacking, we shift each velocity on the 2D grid to the centre of the box, applying periodic boundary conditions, and rotate the velocity field around the central grid such that the central velocity vector points horizontally from the left to the right. Each of these shifted-rotated velocity field is called $\vec v_i'(x,y)$, where $i=1, 2...200^2$. We then average all the $200^2$ fields of $\vec v_i'(x,y)$, yielding the 2D plot shown on the left-hand side of Fig.~\ref{v-correlation-2D}. 

From Fig.~\ref{v-correlation-2D}, we can see a clear dipolar pattern for the velocity field. Velocities diverge from an average under-density on the left, and converge into an average over-density on the right. This resembles the pattern of an electric field generated by an electric dipole, but the coherent pattern is on a scale of hundreds of Mpc. This is successfully predicted from linear theory using Eq.~(\ref{eq:v2D}).

Figure~\ref{Tisw-2D} shows the corresponding ISW dipole from the velocity fields presented in Fig.~\ref{v-correlation-2D}. We sum the contributions along the $z$ direction for a slab of thickness 400$\mpcoh$. The dipole of the lensing convergence presented in Figure~\ref{kappa-2D} is estimated in the same way. From Figure~\ref{Tisw-2D}, we can see a strong coherent temperature dipole on scales of hundreds of Mpc, with an amplitude of a few $\mu$K. This is perfectly as expected from linear theory (left), and can be understood from the velocity field in Fig.~\ref{v-correlation-2D}. The convergence flow corresponds to an over-dense region, and thus its underlying gravitational potential is negative -- a gravitational well. Due to the effect of dark energy, the potential well is becoming shallower, causing CMB photons to gain energy during their traversal and thus generate a CMB hot spot. The divergence flow corresponds to an under-dense region, and thus its underlying gravitational potential is positive -- a gravitational hill. Due to the effect of dark energy, the potential hill is becoming flatter, causing CMB photons to lose energy during their traversal, and thus generating a CMB cold spot. The total effect is a temperature dipole along the horizontal direction. 

\section{Measuring velocity correlations from simulations}
\label{appB}
Following Eqs. 2a, 2b, 3a, 3b of \cite{Gorski1988A}, or Eqs. 10-13 of \cite{Turner2021}, we can measure the two velocity correlation functions in simulations with the following estimators:
\begin{eqnarray}
 \Psi_{\perp} & = &\frac{\Psi_2A_1-\Psi_1A_2}{A_1-A_2}, \\
 \Psi_{\parallel}&=&\frac{(\Psi_2 -\Psi_1) + A_2\Psi_1-A_1\Psi_2}{A_2-A_1},
\end{eqnarray}
where
\begin{eqnarray}
\Psi_1&=&\frac{\sum_{\rm pairs}u_1u_2\cos{\theta_{12}}}{\sum_{\rm pairs}\cos^2{\theta_{12}}}, \\
\Psi_2 &=& \frac{\sum_{\rm pairs} u_1u_2\cos{\theta_1}\cos{\theta_2}}{\sum_{\rm pairs}\cos{\theta_{12}}\cos{\theta_1}\cos{\theta_2} }, 
\end{eqnarray}
\begin{eqnarray}
A_1&=&\frac{\sum_{\rm pairs}\cos{\theta_1}\cos{\theta_2}\cos{\theta_{12}}}{\sum_{\rm pairs}\cos^2{\theta_{12}}}, \\
A_2 &=& \frac{\sum_{\rm pairs} \cos^2{\theta_1}\cos^2{\theta_2}}{\sum_{\rm pairs}\cos{\theta_{12}}\cos{\theta_1}\cos{\theta_2} }, 
\end{eqnarray}
and 
$u={\bf v} \cdot {\bf \hat r}$ is the radial velocity, and ${\bf r}=r{\bf \hat r} = ({\bf r_2}-{\bf r_1})$ is the vector connecting the two points. $\cos{\theta_1}={\bf \hat r_1\cdot \hat r}$, $\cos{\theta_2}={\bf \hat r_2\cdot \hat r}$ and $\cos{\theta_{12}}={\bf \hat r_1\cdot \hat r_2}$.

\bibliographystyle{mnras}
\input{references.bbl}
\end{document}